\documentclass[journal]{IEEEtran}
\usepackage{graphicx}
\usepackage{subfigure}
\usepackage{subfloat}
\usepackage{booktabs}
\usepackage{picinpar}
\usepackage{amsfonts,amssymb}
\usepackage{mathrsfs}
\usepackage{amsmath}
\usepackage[numbers,sort&compress]{natbib}

\begin{document}
%
\title{\huge \bf Robust Coding of Encrypted Images via Structural Matrix}

\author{Yushu Zhang,
        Kwok-Wo Wong,~\IEEEmembership{Senior Member,~IEEE},
        Leo Yu Zhang,
        Di Xiao,~\IEEEmembership{Member,~IEEE},

\thanks{Y. Zhang, D. Xiao are with College of Computer Science, Chongqing University, Chongqing, 400044 China (e-mail: yushuboshi@163.com; xiaodi\_cqu@hotmail.com) }
\thanks{K.W. Wong and L. Zhang are with Department of Electronic Engineering, City University of Hong Kong, Kowloon, Hong Kong (e-mail: itkwwong@cityu.edu.hk; leocityu@gmail.com)}}


\vspace{-0.35in}

\markboth{}
{Shell \MakeLowercase{\textit{et al.}}: Bare Demo of IEEEtran.cls for Journals}
%

\maketitle


\begin{abstract}
The robust coding of natural images and the effective compression of encrypted images have been studied individually in recent years. However, little work has been done in the robust coding of encrypted images. The existing results in these two individual research areas cannot be combined directly for the robust coding of encrypted images. This is because the robust coding of natural images relies on the elimination of spatial correlations using sparse transforms such as discrete wavelet transform (DWT), which is ineffective to encrypted images due to the weak correlation between encrypted pixels. Moreover, the compression of encrypted images always generates code streams with different significance. If one or more such streams are lost, the quality of the reconstructed images may drop substantially or decoding error may exist, which violates the goal of robust coding of encrypted images. In this work, we intend to design a robust coder, based on compressive sensing with \emph{structurally random matrix}, for encrypted images over packet transmission networks. The proposed coder can be applied in the scenario that Alice needs a semi-trusted channel provider Charlie to encode and transmit the encrypted image to Bob. In particular, Alice first encrypts an image using globally random permutation and then sends the encrypted image to Charlie who samples the encrypted image using a \emph{structural matrix}. Through an imperfect channel with packet loss, Bob receives the compressive measurements and reconstructs the original image by joint decryption and decoding. Experimental results show that the proposed coder can be considered as an efficient multiple description coder with a number of descriptions against packet loss.
\end{abstract}


\IEEEpeerreviewmaketitle

\vspace{0.2in}
\section{Introduction}

\IEEEPARstart{T}{}HE traditional approach of transmitting an image via a communication channel is to perform compression preceding encryption at the sender side; and to decrypt the cipher-image followed by decompression at the receiving side. However, consider a particular scenario in which Alice needs to transmit an image to Bob but wants to keep the image confidential to an untrusted channel provider Charlie. This implies that Alice should encrypt the image moderately and Charlie has to compress the encrypted image without any knowledge of the cryptographic key. At the receiving side, Bob performs both decompression and decryption to reconstruct the original image.

Some works for compressing encrypted images have been reported in recent years. A scheme for compressing encrypted images using a 2-D source model and LDPC codes was developed in \cite{schonberg2006compression}. It is based on the finding that encrypted data are as compressible as unencrypted ones by considering the problem as distributed source coding. The lossless compression of encrypted grayscale and color images has been presented in \cite{lazzeretti2008lossless}, by decomposing the image pixels into bit-planes. By applying the approach of \cite{johnson2004compressing} to the prediction error domain, a better lossless compression performance on the encrypted grayscale and color images is achieved \cite{anil2008distributed}. A progressive compression approach for processing an encrypted image has been suggested, in which the decoder needs to study the local statistics of a low-resolution image and then decodes the next resolution level \cite{liu2010efficient}. Meanwhile, the lossy compression of encrypted images was also studied to achieve higher compression ratios \cite{schonberg2008toward, johnson2004compressing, zhang2011lossy, zhang2012scalable, zhang2013compression, zhou2013designing}. For example, based on the results of \cite{johnson2004compressing}, a practical model for compressing encrypted binary image has been developed in \cite{schonberg2008toward}. Zhang proposed a novel scheme for the lossy compression of an encrypted image at a flexible compression ratio \cite{zhang2011lossy}, in which a pseudorandom permutation is used to encrypt the plain-image. Making use of the process of masking the original pixel values by a modulo-256 addition with pseudorandom numbers, Zhang \emph{et al.} further proposed a scheme for the scalable coding of encrypted images \cite{zhang2012scalable}. In \cite{zhang2013compression}, the compression is performed on an encrypted image with multi-layer decomposition. Zhou \emph{et al.} designed an efficient encryption-then-compression scheme for images via error clustering, in which both lossless and lossy compressions were considered \cite{zhou2013designing}. The above-mentioned approaches of compressing encrypted images are not suitable for high packet loss transmission in non-feedback systems, since the resultant coded streams have substantially unequal importance such that the loss of some codewords may cause severe error propagation and results in unsatisfactory decoded result.

Multiple description coding is a common approach to deal with packet loss during transmission. In general, a multiple description coder generates two or more sub-streams referred to as descriptions. The packets of each description are transmitted over multiple disjoint paths. After receiving each description, the decoder is able to perform a low-quality reconstruction. If all the descriptions have been received, the reconstruction quality is the best. Such a protocol allows a channel with network congestion or packet loss to perform the decoding at the expense of reconstruction quality. Multiple description coding of natural images has been extensively studied in \cite{servetto2000multiple, yang2007robust, li2011balanced, deng2012robust}, where spatial correlations are often eliminated by using sparse transforms like DWT. However, they are not suitable for encrypted images since sparse transforms are nearly ineffective on encrypted images due to the low correlation between the pixels. A multiple description coder especially designed for encrypted images is rarely reported so far.

Consider the scenario that Alice needs the semi-trusted channel coder Charlie to transmit an encrypted image to Bob. When a high packet loss is encountered in the channel between Charlie and Bob, Charlie should first encode the encrypted image for error control. This motivates us to explore a multiple description coder aiming at the robust coding of encrypted images. In this work, we design such a coder based on compressive sensing (CS) with a structurally random matrix (SRM). The proposed coder is comprised of three parts: permutation-based encryption by Alice, encoding using structural matrix (SM) by Charlie, and joint decryption and decoding by Bob. In particular, Alice first encrypts an image using globally random permutation and then sends the encrypted image to the semi-trusted channel encoder Charlie who samples the encrypted image using a structural matrix. Through a channel with high packet loss, Bob receives the compressive measurements and reconstructs the original image by joint decryption and decoding. Moreover, we discuss the relationship between our approach and existing algorithms and describe two other cryptographic applications of SRM. In the performance evaluation, we explore the relationship between packet loss rate and sampling rate and then introduce a feasible quantization approach to the compressive measurements of encrypted images. Finally, we investigate the robustness of the proposed coder at different parameter settings. It is verified that the proposed coder can be regarded as an efficient multiple description coder with a number of descriptions against packet loss.

The rest of this paper is organized as follows. Section II is a brief review of the theory of CS using SRM. In Section III, the robust coding of encrypted images based on CS with SM is proposed. Further discussions can be found in Section IV while the performance evaluation is reported in Section V. Finally, we conclude the paper with some remarks in Section VI.

\vspace{-0.05in}
\section{Compressive Sensing by Structurally Random Matrix}
The fundamental Shannon/Nyquist sampling theory is widely-accepted as the keystone in signal acquisition and reconstruction. It governs the sampling process from the perspective of signal bandwidth. Nevertheless, the number of required measurements can be so large that the storage becomes unbearable and the acquisition time can be very long. Compressive sensing \cite{candes2006robust, donoho2006compressed} is a new sampling theory which allows the exact recovery of a sparse signal from a few linear projections lower than the Nyquist rate. The underlying property of CS is the sparsity of interest. A signal ${\bf{x}}$ of length $N$ is said to be $K$-sparse or compressible if it can be well approximated using only $K \ll N$ coefficients over some sparsifying basis ${\bf{\Psi }}$ as follows

\begin{equation}
{\bf{x = \Psi s}},
\end{equation}
where ${\bf{s}}$ is the transform coefficient vector that contains at most $K$ significant nonzero entries. Compressive sensing theory indicates that ${\bf{x}}$ can be acquired by the following random measurement

\begin{equation}
{\bf{y = \Phi x}},
\end{equation}
where ${\bf{\Phi }}$ is a $M \times N$ ($M < N$) random measurement matrix and ${\bf{y}}$ represents the measurement coefficient vector. ${\bf{x}}$ can be faithfully recovered from only $M = {\rm \mathcal{O}}\left( {K\log N} \right)$ measurements through ${l_1}$-minimization

\begin{equation}
\min {\kern 1pt} {\kern 1pt} {\left\| {\bf{s}} \right\|_1}\; \textrm{s.t.} \;{\bf{y = \Phi \Psi s}},\;\\
\end{equation}
where the measurement matrix ${\bf{\Phi }}$ should be highly incoherent with the sparsifying basis ${\bf{\Psi }}$.

The design of an efficient measurement matrix is still a big challenge in CS. Do \emph{et al.} \cite{do2012fast} introduced a fast and efficient measurement matrix for practical CS. The matrix is called a structurally random matrix (SRM), which, in many aspects, outperforms the existing popular sensing matrices such as Gaussian, Bernoulli and Fourier matrices \cite{candes2006near, mendelson2008uniform, candes2007sparsity}. Gaussian and Bernoulli matrices require high computation complexity and huge memory buffering due to their completely unstructured nature while Fourier matrix works well only if the sparsifying basis is an identity matrix. Do \emph{et al.} also pointed out that SRM possesses the following features: optimal or near-optimal sensing performance; universality; low complexity; hardware/optical implementation friendless. In particular, it is defined as a product of three matrices

\begin{equation}
{\bf{\Phi }} = \sqrt {\frac{N}{M}} {\bf{DFR}}
\end{equation}
where ${\bf{R}} \in {\mathbb{R}^{N \times N}}$ is either a uniform random permutation matrix or a diagonal random matrix whose diagonal entries are Bernoulli random variables. ${\bf{F}} \in {\mathbb{R}^{N \times N}}$ represents an orthonormal matrix that is selected among popular fast computable transforms such as Fast Fourier Transform (FFT), Discrete Cosine Transform (DCT) and Walsh-Hadamard Transform (WHT). ${\bf{D}} \in {\mathbb{R}^{N \times N}}$ is a subsampling operator selecting a random subset of rows of the matrix ${\bf{FR}}$. Interested readers can refer to \cite{do2012fast} for more details on SRM.

\vspace{-0.05in}
\section{Robust Coding of Encrypted Image via Structural Matrix}

Compressing encrypted images is a big challenge due to the fact that an effective encryption algorithm must have already removed or lowered the correlation among neighbouring image pixels to increase the entropy. However, classical image compression schemes like JPEG 2000 always make use of the high correlation and non-uniformity of image pixels. Some lightweight encryption techniques only permute the pixels or mask the pixel values by a keystream. As a result, the encrypted image may still be compressed to certain extent by leveraging some particular coding techniques \cite{johnson2004compressing, schonberg2006compression, lazzeretti2008lossless, anil2008distributed, liu2010efficient, schonberg2008toward, zhang2011lossy, zhang2012scalable, zhang2013compression, zhou2013designing}. The lightweight encryption schemes are usually not secure enough, but they are employed in some specific application scenarios. The proposed scheme does not aim at improving the compression performance on encrypted images but focuses on designing a robust coder for the transmission of encrypted images over a channel with high packet loss rate.

The proposed coder is based on SRM. The basic idea is to split the measurement matrix ${\bf{\Phi }} = \sqrt {{N \mathord{\left/
{\vphantom {N M}} \right. \kern-\nulldelimiterspace} M}} {\bf{DFR}}$ in (4) into two matrices: the matrix ${\bf{R}}$ and the matrix $\sqrt {{N \mathord{\left/
{\vphantom {N M}} \right.  \kern-\nulldelimiterspace} M}} {\bf{DF}}$. ${\bf{R}}$ is a random permutation matrix which can serve as a lightweight encryption tool while $\sqrt {{N \mathord{\left/
{\vphantom {N M}} \right.  \kern-\nulldelimiterspace} M}} {\bf{DF}}$ can be considered as a new measurement matrix in the proposed coder. First, Alice encrypts an image using ${\bf{R}}$ and then sends the encrypted image to the channel coder Charlie who samples the encrypted image using $\sqrt {{N \mathord{\left/ {\vphantom {N M}} \right.  \kern-\nulldelimiterspace} M}} {\bf{DF}}$. Through a high packet loss channel, Bob receives the compressive measurements and reconstructs the original image by joint decryption and decoding using $\sqrt {{N \mathord{\left/ {\vphantom {N M}} \right. \kern-\nulldelimiterspace} M}} {\bf{DFR}}$, as illustrated in Fig. 1. The random permutation ${\bf{R}}$ is constructed from a secret seed known to both Alice and Bob.

The robust coding of encrypted images by structural matrices is composed of three steps: permutation-based encryption by Alice, encoding using structural matrix by Charlie, and joint decryption and decoding by Bob.

\begin{figure}[th]\centering
 \includegraphics[width=8 cm]{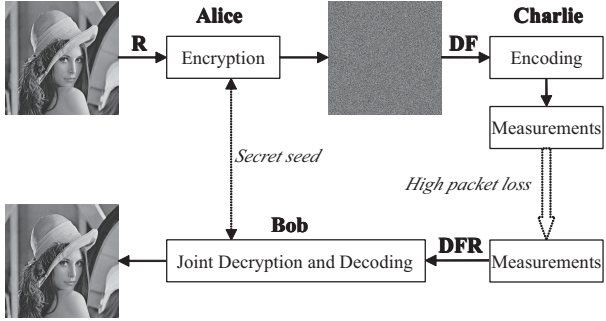}\\
  \caption{A block diagram of the proposed coder.}
  \label{fig1}
\end{figure}

\subsection{Permutation-based Encryption by Alice}

The encrypted image is obtained by applying random spatial permutation on the image. Alice converts the original image ${\bf{X}}$ of size ${N_1} \times {N_2}$ into a vector ${\bf{x}}$ with length $N = {N_1} \times {N_2}$. Then she encrypts ${\bf{x}}$ to the cipher sequence ${{\bf{x}}_{en}}$ by applying a random permutation matrix ${\bf{R}} \in {\mathbb{R}^{N \times N}}$, governed by

\begin{equation}
{{\bf{x}}_{en}} = {\bf{Rx}}.
\end{equation}
${{\bf{x}}_{en}}$ is rearranged into a 2-D cipher image ${{\bf{X}}_{en}}$, which is then sent to Charlie who obtains the encrypted sequence ${{\bf{x}}_{en}}$ by arranging ${{\bf{X}}_{en}}$. The conversion between vector and matrix is known to both Alice and Charlie. The random permutation matrix ${\bf{R}}$ is a binary matrix in which each row or column has exactly one 1 and the rest are all zero. It is generated by a pseudo-random generator with initial random seed shared between Alice and Bob. The reader may refer to \cite{mitra2006new, zhang2014image} for more illustrations on the encryption methods based on permutation matrix. It should be noticed that permutation-based decryption is performed by multiplying the cipher image with the inverse permutation matrix. Interestingly, it is not necessary to invert the matrix since the inverse matrix is obtained by transposing the permutation matrix itself, i.e., ${{\bf{R}}^{ - 1}} = {{\bf{R}}^T}$. The key space is $N!$ so that it is not likely for Charlie to launch a brute force search when $N$ is sufficiently large. Permutation-based encryption cannot hide the statistical information of the original image due to its unaltered histogram. In spite of this, it can still be employed in applications where high secrecy is not a must.

\subsection{Encoding using Structural Matrix by Charlie}

After the encrypted image has been received, Charlie constructs a special measurement matrix to sample it. This matrix is tailored to the encrypted image and is called structural matrix (SM). It is governed by

\begin{equation}
{\bf{A}} = \sqrt {\frac{N}{M}} {\bf{DF}},
\end{equation}
where ${\bf{D}}$ and ${\bf{F}}$ are as described in (4). Encoding using SM is expressed as

\begin{equation}
{\bf{y = A}}{{\bf{x}}_{en}}.
\end{equation}
Obviously, SM is derived from SRM due to the fact that ${\bf{y = A}}{{\bf{x}}_{en}} = \sqrt {{N \mathord{\left/ {\vphantom {N M}} \right. \kern-\nulldelimiterspace} M}} {\bf{DF}}{{\bf{x}}_{en}} = \sqrt {{N \mathord{\left/ {\vphantom {N M}} \right. \kern-\nulldelimiterspace} M}} {\bf{DFRx}} = {\bf{\Phi x}}$. The scenario that SM is applied for permuted or encrypted images is the same as that SRM is employed for spatial images. Structural matrix is expediently selected among some popular computable matrices such as FFT, SCT, WHT or their block diagonal versions. The $M$ rows are extracted  at random from SM. These matrices have stable structures like SRM and they outperform Gaussian and Bernoulli matrices in terms of computational complexity and memory requirement. It can be easily inferred that the performance of SM measuring the encrypted image is the same as that of SRM sampling the original image. It has been mathematically proved in \cite{do2012fast} that entries of ${\bf{AR\Psi }}$ asymptotically form a normal distribution $\mathcal{N}\left( {0,{\sigma ^2}} \right)$, where ${\bf{\Psi }}$ is an arbitrary orthonormal matrix and ${\sigma ^2} \le {\rm \mathcal{O}}\left( {\frac{1}{\mathcal{N}}} \right)$, under some mild assumptions: ${\bf{F}}$ is an unit-row matrix whose entries have absolute magnitude in the order of ${\sigma ^2} \le {\rm \mathcal{O}}\left( {\frac{1}{\mathcal{N}}} \right)$ and the sum of entries in each row is equal to zero; ${\bf{\Psi }}$ is an unit-norm column matrix with entries having maximal absolute magnitude in the order of $\mathcal{O}\left( 1 \right)$ and the average sum of entries in each column in the order of ${\sigma ^2} \le {\rm \mathcal{O}}\left( {\frac{1}{\mathcal{N}}} \right)$. The entries in each row of ${\bf{F}}$ and each column of ${\bf{\Psi }}$ are not all equal. Do \emph{et al.} also found that SRM supports block-based models with high incoherence between ${\bf{FR}}$ and ${\bf{\Psi }}$. It should be noticed that the randomization ${\bf{D}}$ can induce a new application scenario, which will be described later.

\subsection{Joint decryption and decoding by Bob}

At the receiving side, Bob obtains the compressive measurements ${\bf{y}}$ and applies joint decryption and decoding to recover the original image using the following algorithm:

\begin{equation}
\min {\kern 1pt} {\kern 1pt} {\left\| {\bf{s}} \right\|_1}\; \textrm{s.t.} \;{\bf{y = AR\Psi s}} = \sqrt {\frac{N}{M}} {\bf{DFR\Psi s}}\;\\
\end{equation}
As a result, ${\bf{x = \Psi s}}$.  The recovery criterion has been stated in \cite{do2012fast}: with a probability of at least $1 - \delta $, the sensing framework using SRM can exactly recover $K$-sparse signals if $M \ge {\rm \mathcal{O}}\left( {\frac{N}{B}K{{\log }^2}\frac{N}{\delta }} \right)$, where $B$ is the block size. Theoretically, this guarantees the capability of SM in encoding the encrypted image.

\vspace{-0.05in}
\section{Further Discussions}

In some references \cite{han2010image, wu2011multivariate, zhang2012image, gao2013image}, CS was applied for natural image coding but this is not an appropriate approach in terms of compression efficiency \cite{ goyal2008compressive}. Nevertheless, in view of the robustness property of multiple description coder, CS can be a good candidate \cite{gao2010robust, liu2011new, deng2012robust}. A representative work was presented by Deng \emph{et al.} in \cite{deng2012robust}, in which the compressive measurements can be viewed as a number of descriptions mainly because of their \emph{democracy} properties. If the measurement matrix follows the Gaussian distribution, each CS measurement possesses a similar amount of information of the original signal \cite{davenport2009simple}. Specifically, the sampling is performed on the frequency coefficients generated by two-dimensional DWT and at the decoding side, two different recovery algorithms are developed for the low-frequency and high-frequency subbands, respectively, by fully exploiting the intra-scale and inter-scale correlation of multiscale DWT. Although experimental results showed that this CS-based codec is much more robust for lossy channels in comparison with existing CS-based coding schemes \cite{deng2012robust}, it is not suitable for processing encrypted images. This is because the efficiency of sparse transforms like DWT mainly depends on strong correlation between pixels, which must be weakened by the encryption process, even if a lightweight one is employed.

CS-based compression of encrypted image has been explored in only two references \cite{kumar2009lossy, zhang2011compressing}, both of which aimed at the linear transformation encryption operations. Both coders adopt the block-to-block structure which possesses a straightforward advantage, i.e., parallel CS encoding and decoding. Unfortunately, such a block encryption manner suffers from three drawbacks. Firstly, individual block operation makes the cipher more insecure than global image transform. In order to enhance the security, different blocks may be endowed with different keys and more keys need to be transmitted. Secondly, a plain image is divided into a number of non-overlapping blocks having different statistical features and unequal significance. When these blocks are individually sampled, the measurements have unequal significance. As a result, both coders cannot be considered as efficient multiple description coders. Thirdly, blocking artifact cannot be avoided. In addition, a random matrix is chosen as the measurement matrix. In practical sensing applications, this is costly as very high computational complexity and huge memory buffering are required due to the completely unstructured nature of the matrix \cite{candes2007sparsity}. The proposed coder does not suffer from the above drawbacks. Global permutation is a common lightweight image encryption technique which is more secure than individual block permutation. The random permutation ${\bf{R}}$ relocates all the pixels globally. It destroys the image structure and converts a meaningful image into one look like white noise \cite{do2012fast}. The structural matrix ${\bf{A}}$ in sampling the permuted image supports block processing, meaning that parallel CS encoding can be applied. ${\bf{R}}$ disperses the energy of the whole image and ${\bf{F}}$ further spreads the energy over all the measurements. Consequently, the sampled measurements obtained by SM roughly have the same significance. The proposed coder is a multiple description coder with a number of descriptions whose capability in resisting against packet loss is verified in the next section. There is no blocking artifact as a unified decoder is used to reconstruct the whole image. Compared with random matrix, SM facilitates fast computation and low-complexity electronic or optical implementation.

It is worth mentioning that SRM also induces two other applications related to coding and encryption due to the randomness of ${\bf{R}}$ and ${\bf{D}}$. The first application is illustrated in Fig. 2(a). Alice still permutes the image with ${\bf{R}}$ while Charlie can further encrypt the permuted image with ${\bf{DF}}$. This is because the matrix ${\bf{D}}$ is a random selection operation which can serve as a secret key shared between Charlie and Bob. Another application is the direct encryption by Alice using ${\bf{DFR}}$, as shown in Fig. 2(b). Both applications can be considered as joint coding and encryption schemes. The size of the key space due to ${\bf{D}}$ is given by the combinatorial number $\left( \begin{array}{l}M\\N\end{array} \right)$. It seems that the current size of key space upgraded as $N! + \left( \begin{array}{l}M\\N\end{array} \right)$ is sufficiently large to resist brute-force attack. Unfortunately, the encryption schemes based on CS with SRM is probably insecure against some potential attacks such as known-plaintext attack and chosen-plaintext attack due to its linearity \cite{rachlin2008secrecy}. As a consequence, the security level of CS needs to be analyzed. For example, a low-complexity multiclass encryption scheme has been designed in \cite{cambareri2013low1, cambareri2013low2}, which possesses strong resistance against known-plaintext attacks.

\begin{figure}[th]
\centering
\subfigure[]{
\begin{minipage}[t]{0.85\linewidth}
\includegraphics[width=\textwidth]{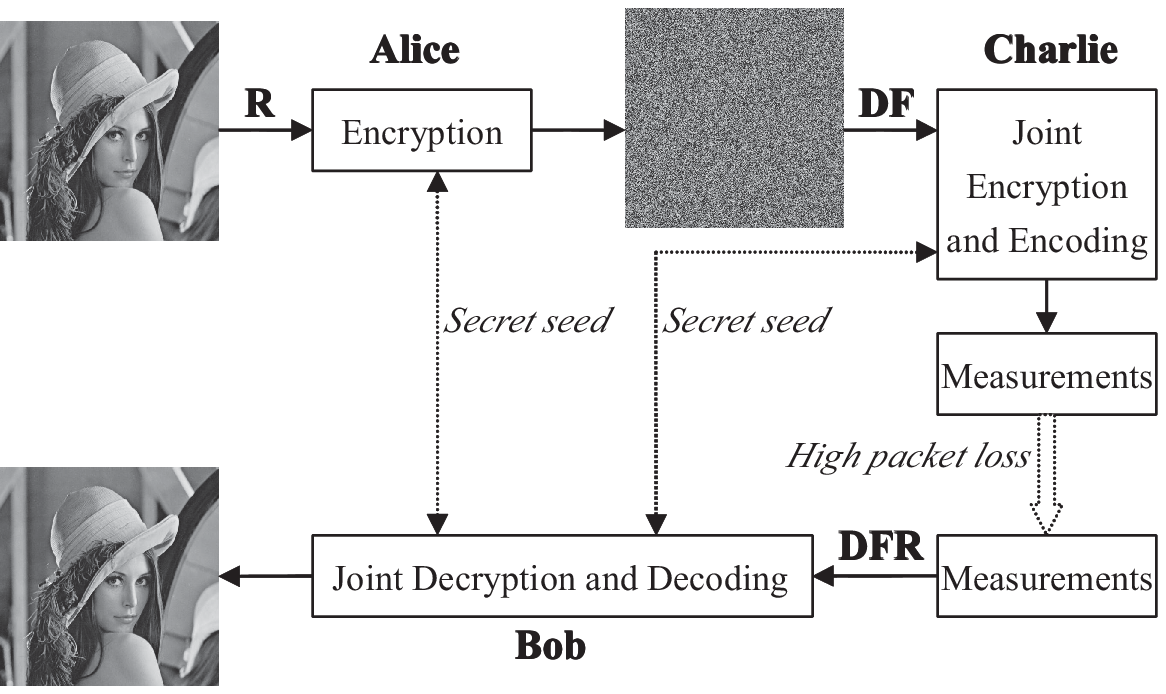}
\end{minipage}}\\
\subfigure[]{
\begin{minipage}[t]{0.85\linewidth}
\includegraphics[width=\textwidth]{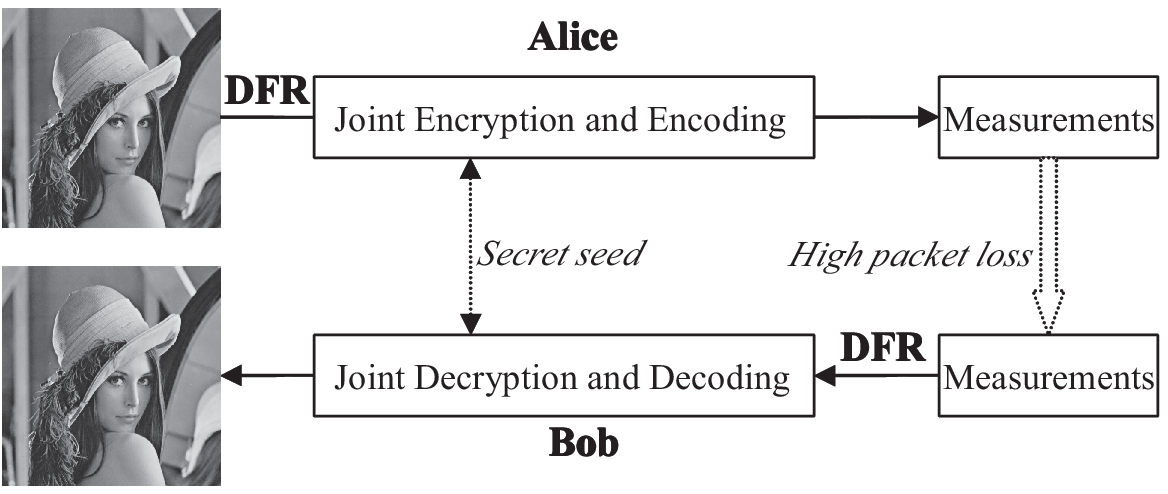}
\end{minipage}}
\caption{\small Two other applications of SRM.}
\label{fig2}
\end{figure}

\begin{figure}[th]
\centering
\subfigure[]{
\begin{minipage}[t]{0.9\linewidth}
\includegraphics[width=\textwidth]{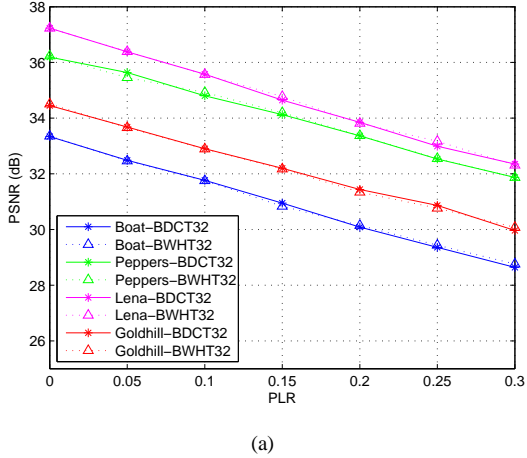}
\end{minipage}}\\
\subfigure[]{
\begin{minipage}[t]{0.9\linewidth}
\includegraphics[width=\textwidth]{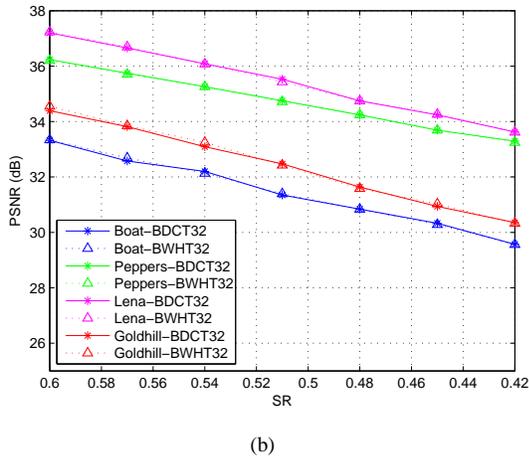}
\end{minipage}}
\caption{\small PSNRs of the reconstructed images with respect to (a) PLR; (b) SR.}
\label{fig3}
\end{figure}

\begin{figure*}[th]
\centering
\subfigure[]{
\begin{minipage}[t]{0.22\linewidth}
\includegraphics[width=\textwidth]{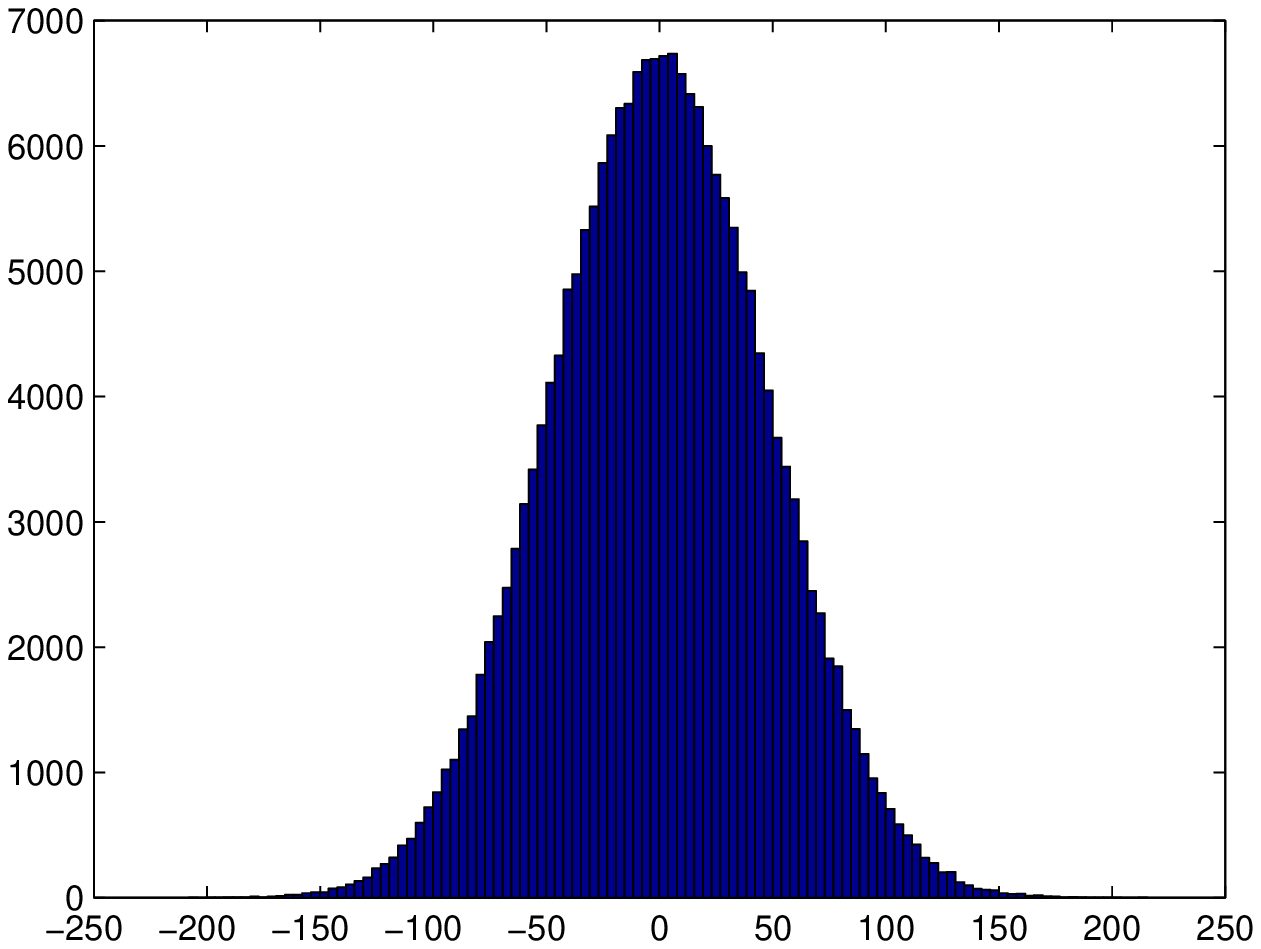}
\end{minipage}}
\subfigure[]{
\begin{minipage}[t]{0.22\linewidth}
\includegraphics[width=\textwidth]{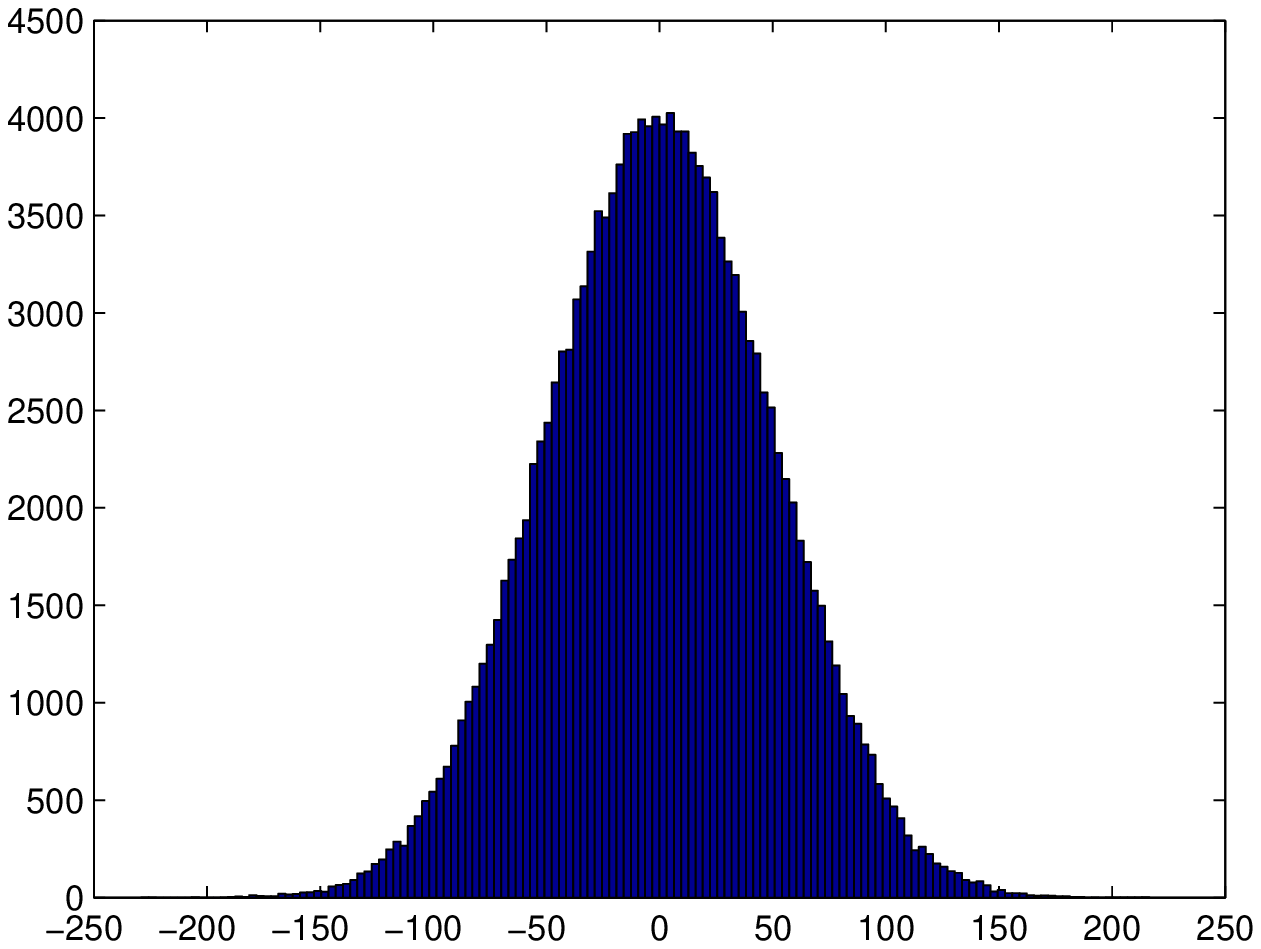}
\end{minipage}}
\subfigure[]{
\begin{minipage}[t]{0.22\linewidth}
\includegraphics[width=\textwidth]{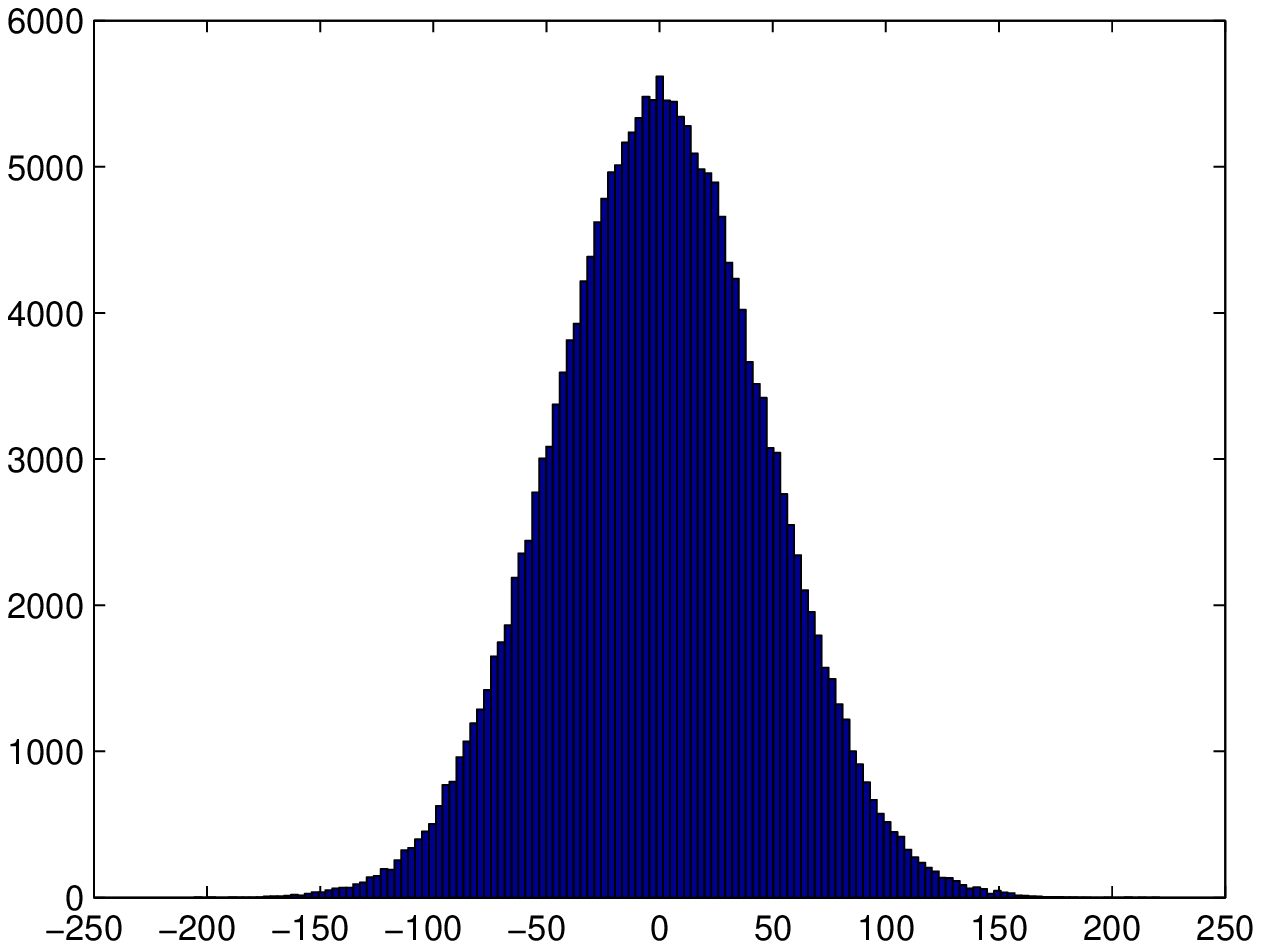}
\end{minipage}}
\subfigure[]{
\begin{minipage}[t]{0.22\linewidth}
\includegraphics[width=\textwidth]{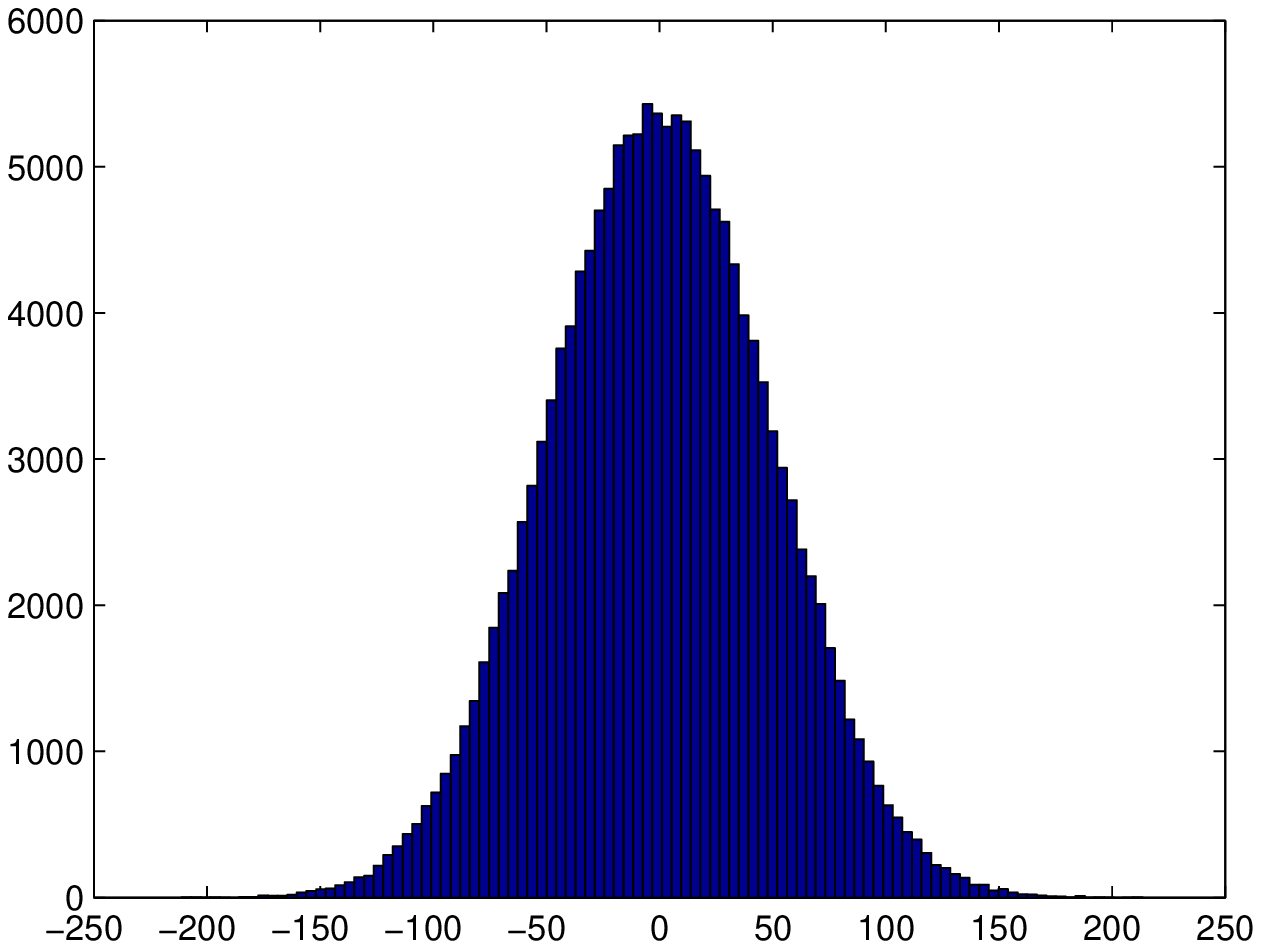}
\end{minipage}}
\caption{\small Histograms of the encoded images for the cases: (a) Lena, SR=0.8, BDCT32, PLR=0.05; (b) Peppers, SR=0.6, BWHT32, PLR=0.10; (c) Boat, SR=0.8, BDCT32, PLR=0.15; (d) Goldhill, SR=0.6, BWHT32, PLR=0.20.}
\label{fig4}
\end{figure*}

\begin{figure}[th]
\centering
\subfigure[]{
\begin{minipage}[t]{0.9\linewidth}
\includegraphics[width=\textwidth]{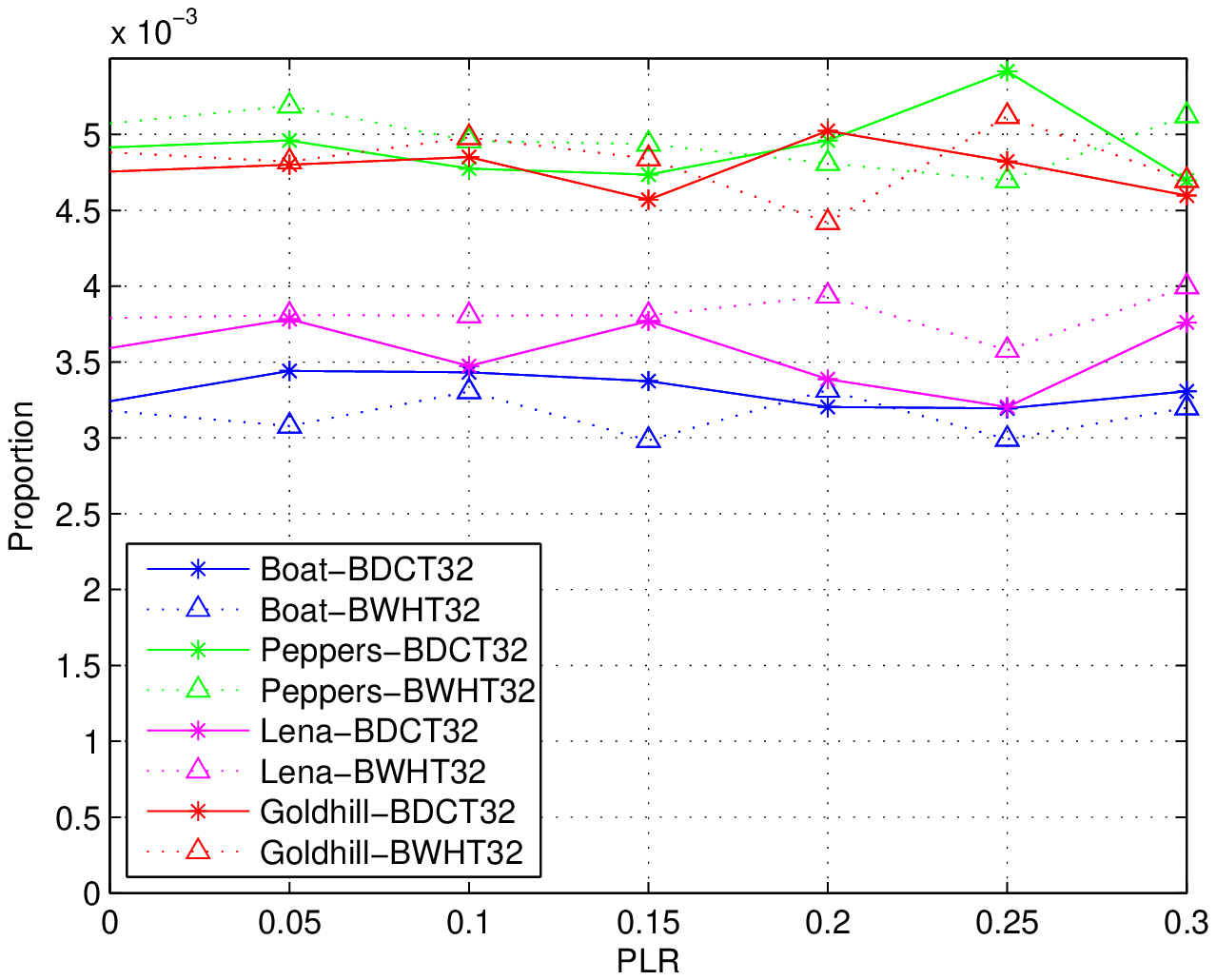}
\end{minipage}}
\subfigure[]{
\begin{minipage}[t]{0.9\linewidth}
\includegraphics[width=\textwidth]{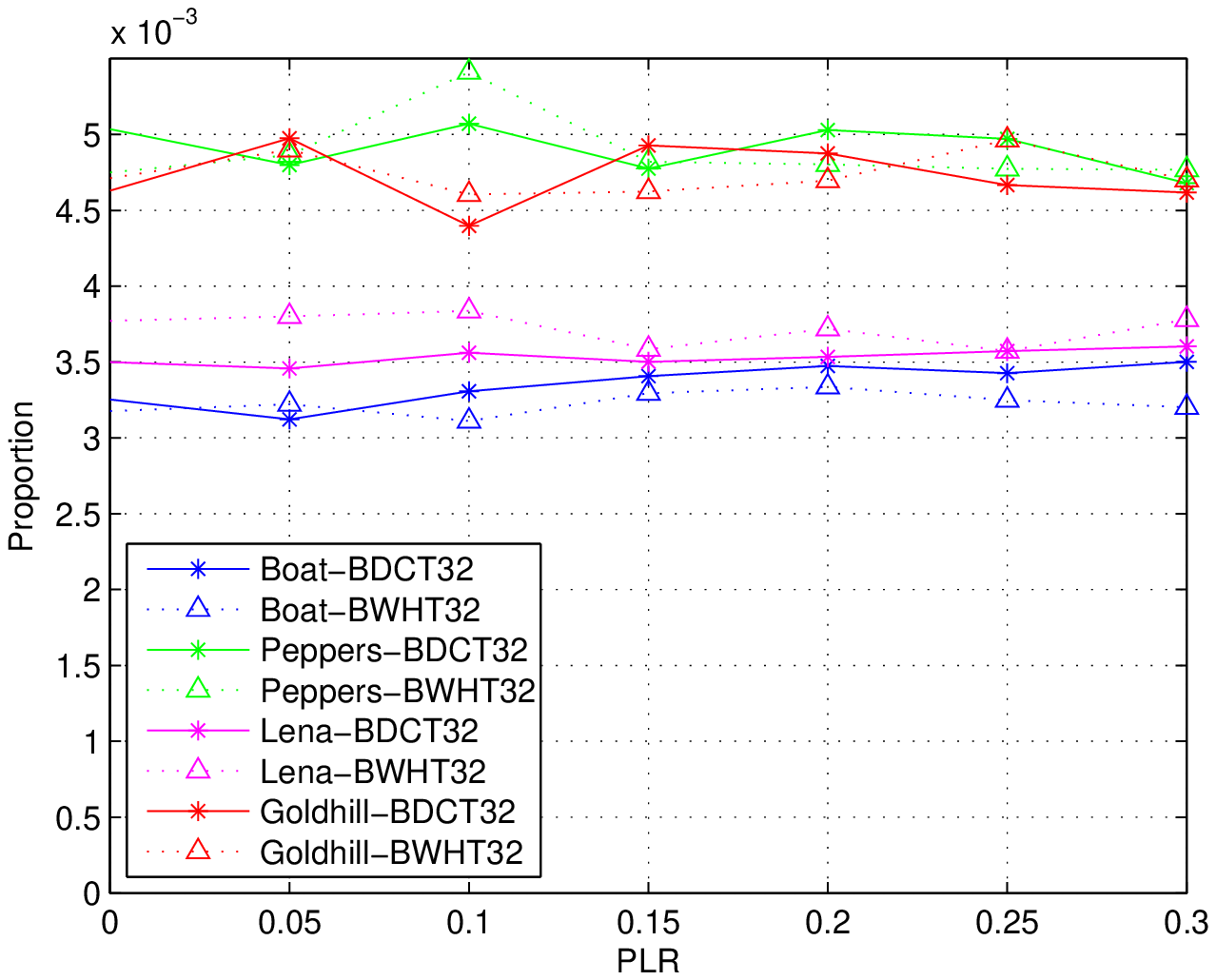}
\end{minipage}}
\caption{\small The values of $\gamma $ versus PLR for (a) SR=0.6; (b) SR=0.8.}
\label{fig5}
\end{figure}

\begin{figure}[th]
\centering
\subfigure[]{
\begin{minipage}[t]{0.9\linewidth}
\includegraphics[width=\textwidth]{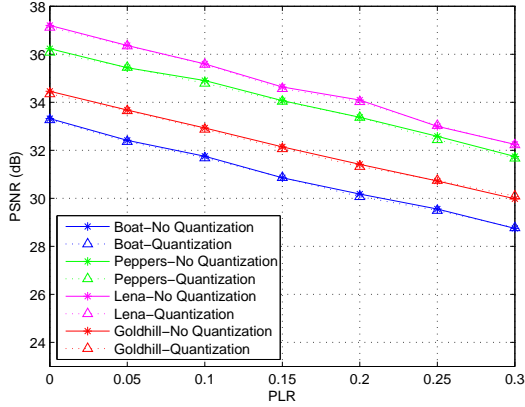}
\end{minipage}}
\subfigure[]{
\begin{minipage}[t]{0.9\linewidth}
\includegraphics[width=\textwidth]{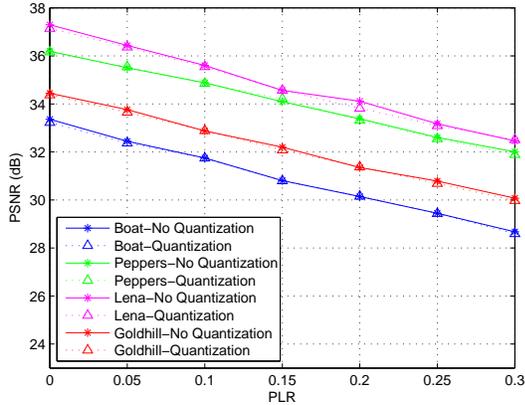}
\end{minipage}}
\caption{\small Rate-Distortion performance of the quantization for (a) SR=0.6, BDCT32; (b) SR=0.6, BWHT32.}
\label{fig6}
\end{figure}

\vspace{-0.05in}
\section{Performance Evaluation}

Our simulation settings are similar to those using SRM \cite{do2012fast}. Four natural images of size 512$\times$512 including \emph{Lena}, \emph{Peppers}, \emph{Boat} and \emph{Goldhill} are used for testing. The sparsifying basis ${\bf{\Psi }}$ is Daubechies 9/7 wavelet transform. The reconstruction algorithm is GRSR in \cite{figueiredo2007gradient}. ${\bf{R}}$ and ${\bf{D}}$ are generated using MATLAB commands and ${\bf{F}}$ is chosen as block diagonal DCT (BDCT) and block diagonal WHT (BWHT). The packet size is set to 100 unless specified. We first explore the relationship between packet loss rate and sampling rate and then describe a feasible quantization approach for the compressive measurements of encrypted images. Finally, the robustness of the proposed coder at different parameter settings is investigated.

\subsection{Relationship between packet loss rate and sampling rate}

The compressive measurements ${\bf{y}}$ of length $M$ can be partitioned, at equal intervals, into a number of packets. Each packet carries a similar amount of information of the original image since all the measurements have roughly equal importance. If a packet contains $m$ measurements, there are $\left\lceil {{M \mathord{\left/ {\vphantom {M m}} \right. \kern-\nulldelimiterspace} m}} \right\rceil $ packets in total. Lost packets always occur randomly and  Bob will update ${\bf{D}}$ according to the received packets. We denote packet loss rate as PLR which can be up to 30\% in real cases \cite{zhao2003understanding}. The sampling rate (SR) is defined as $SR = {M \mathord{\left/ {\vphantom {M N}} \right. \kern-\nulldelimiterspace} N}$. For example, if $M = 157290$ and $m = 100$, then $SR = {{157290} \mathord{\left/ {\vphantom {{157290} {{{512}^2}}}} \right. \kern-\nulldelimiterspace} {{{512}^2}}} = 0.60$ and the number of packets is $\left\lceil {{{157290} \mathord{\left/ {\vphantom {{157290} {100}}} \right. \kern-\nulldelimiterspace} {100}}} \right\rceil  = 1573$. If $PLR = 0.20$, the number of lost packets is $1573 \times 0.2 \buildrel\textstyle.\over= 315$ and the number of received packets is 1258. In other words, Charlie sends ${512^2}$ measurements and Bob receives about 125800 measurements among them. This is similar to the case that the sampling rate is changed to $SR' = {{125800} \mathord{\left/ {\vphantom {{125800} {{{512}^2}}}} \right. \kern-\nulldelimiterspace} {{{512}^2}}} \buildrel\textstyle.\over= 0.48$. In fact, this equivalence is reasonable due to the roughly equal importance of the measurements. This example inspires us a relationship between SR and PLR.

In general, for a given $SR = \alpha $ $\left( {0 < \alpha  < 1} \right)$, $PLR = \beta $ $\left( {0 \le \beta  \le 0.3} \right)$ is basically equivalent to $SR = \alpha \left( {1 - \beta } \right)$. This can be verified in Fig. 3, where BDCT32 and BWHT32, corresponding to the solid line and the dashed line, respectively, mean that each sub-matrix in the diagonal of ${\bf{F}}$ has a size of 32$\times$32. It can be observed that the effects of BDCT and BWHT are consistent since each pair of solid and dashed lines coincides with each other while other conditions are identical. The value of SR is set as $SR{\rm{ = }}0.6$ in Fig. 3(a). $PLR = \beta $ in Fig. 3(a) corresponds to $SR = 0.6 \times \left( {1 - \beta } \right)$ in Fig. 3(b). A comparison between Fig. 3(a) and Fig. 3(b) shows that the former PSNR roughly coincides with the latter one. Both starting points have the same PSNR value, i.e., $PLR = 0$ in Fig. 3(a) and $SR{\rm{ = }}0.6$ in Fig. 3(b). However, with the increase of PLR and the reduction of SR, the PSNR value of the former is sightly lower than that of the latter. There are three factors causing this difference: (i) Weak correlations exist between adjacent measurements. The amount of information of the whole packet containing $m$ successive measurements is gracefully greater than that provided by the $m$ randomly-sampled  measurements; (ii) After packing the measurements, the number of measurements $m'$ in the last packet is less than $m$ as long as $M$ is not divisible by $m$. The last packet will not be lost with high probability $\left( {1 - \beta } \right)$ such that the actual $SR = \alpha {{\left( {m\left( {\left\lceil {{M \mathord{\left/ {\vphantom {M m}} \right. \kern-\nulldelimiterspace} m}} \right\rceil \left( {1 - \beta } \right) - 1} \right) + m'} \right)} \mathord{\left/ {\vphantom {{\left( {m\left( {\left\lceil {{M \mathord{\left/ {\vphantom {M m}} \right. \kern-\nulldelimiterspace} m}} \right\rceil \left( {1 - \beta } \right) - 1} \right) + m'} \right)} M}} \right. \kern-\nulldelimiterspace} M} < \alpha \left( {1 - \beta } \right)$. (iii) The rounding effect of $\left\lceil {{M \mathord{\left/ {\vphantom {M m}} \right. \kern-\nulldelimiterspace} m}} \right\rceil \beta $ possibly results in the actual $PLR = {{round\left( {\left\lceil {{M \mathord{\left/ {\vphantom {M m}} \right. \kern-\nulldelimiterspace} m}} \right\rceil  \cdot \beta } \right)} \mathord{\left/ {\vphantom {{\textrm{round}\left( {\left\lceil {{M \mathord{\left/ {\vphantom {M m}} \right. \kern-\nulldelimiterspace} m}} \right\rceil  \cdot \beta } \right)} {\left\lceil {{M \mathord{\left/ {\vphantom {M m}} \right. \kern-\nulldelimiterspace} m}} \right\rceil }}} \right. \kern-\nulldelimiterspace} {\left\lceil {{M \mathord{\left/ {\vphantom {M m}} \right. \kern-\nulldelimiterspace} m}} \right\rceil }} > \beta $. Revealing such a connection of PLR and SR helps to adjust the SR according to the PLR in real-time transmission. Bob distinguishes the PLR according to the received packets and then feeds back to Charlie who adjusts the SR to guarantee a certain PSNR value for the image received by Bob.

\begin{figure*}[th]
\centering
\subfigure[]{
\begin{minipage}[t]{0.22\linewidth}
\includegraphics[width=\textwidth]{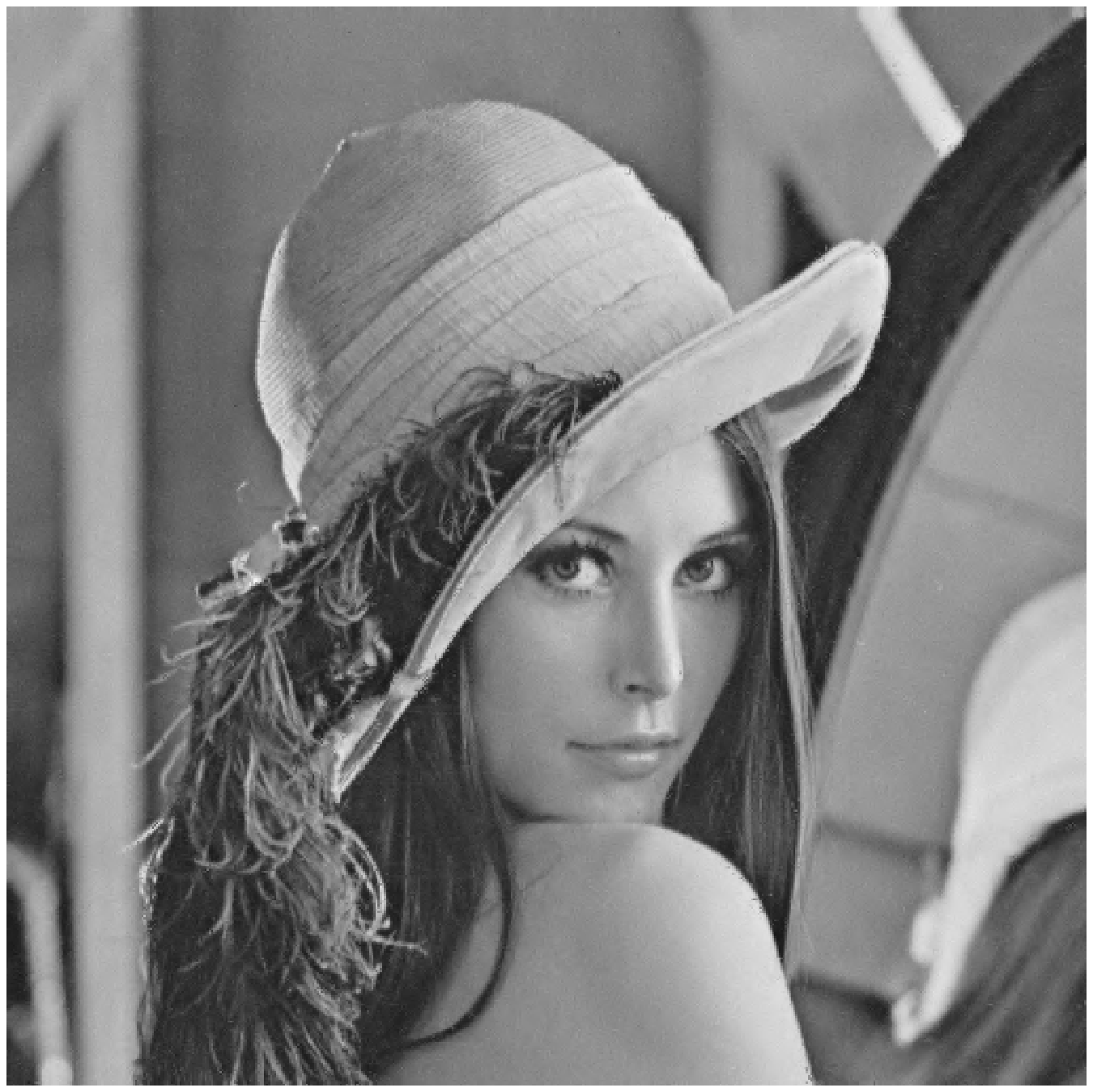}
\end{minipage}}
\subfigure[]{
\begin{minipage}[t]{0.22\linewidth}
\includegraphics[width=\textwidth]{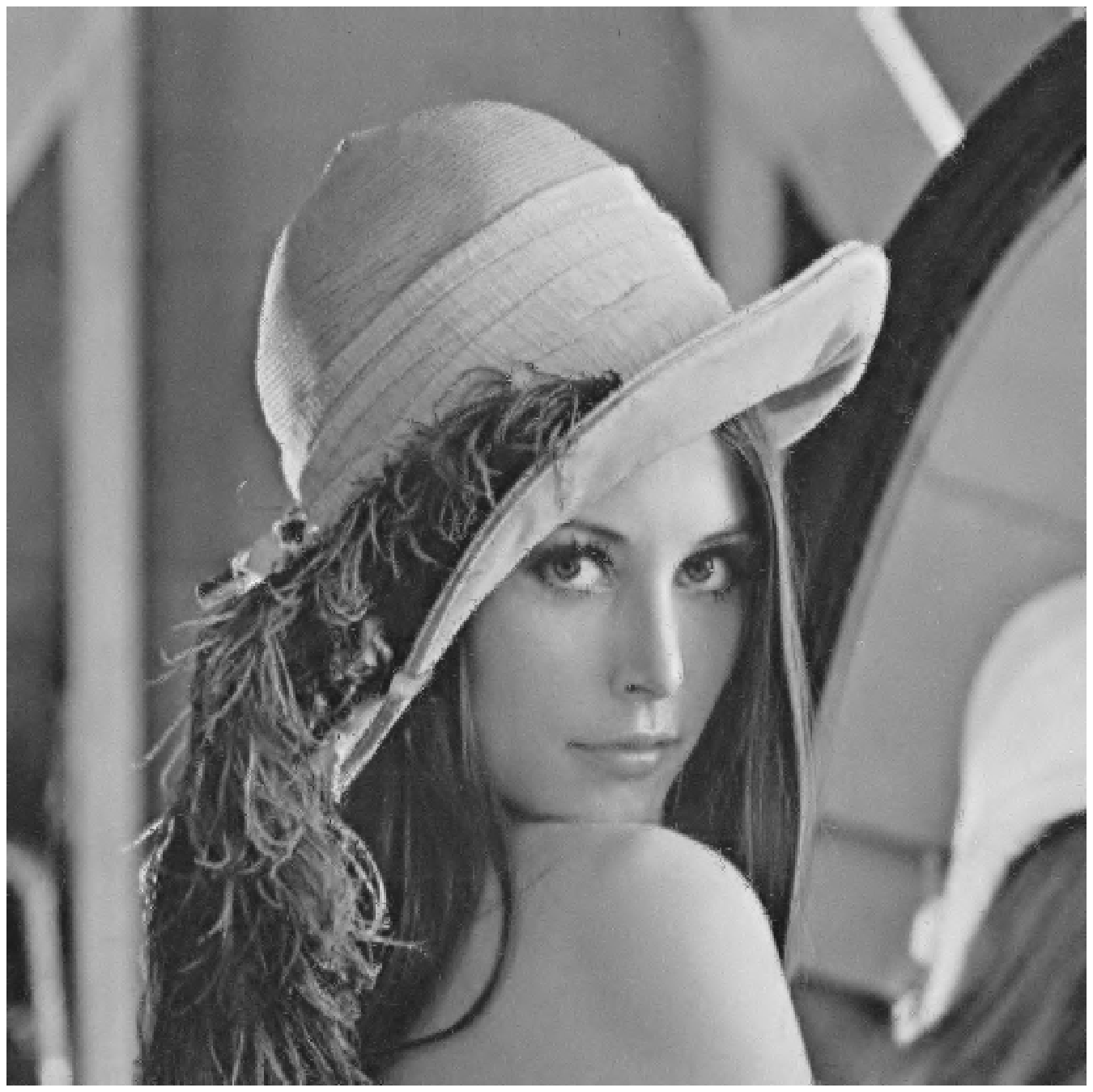}
\end{minipage}}
\subfigure[]{
\begin{minipage}[t]{0.22\linewidth}
\includegraphics[width=\textwidth]{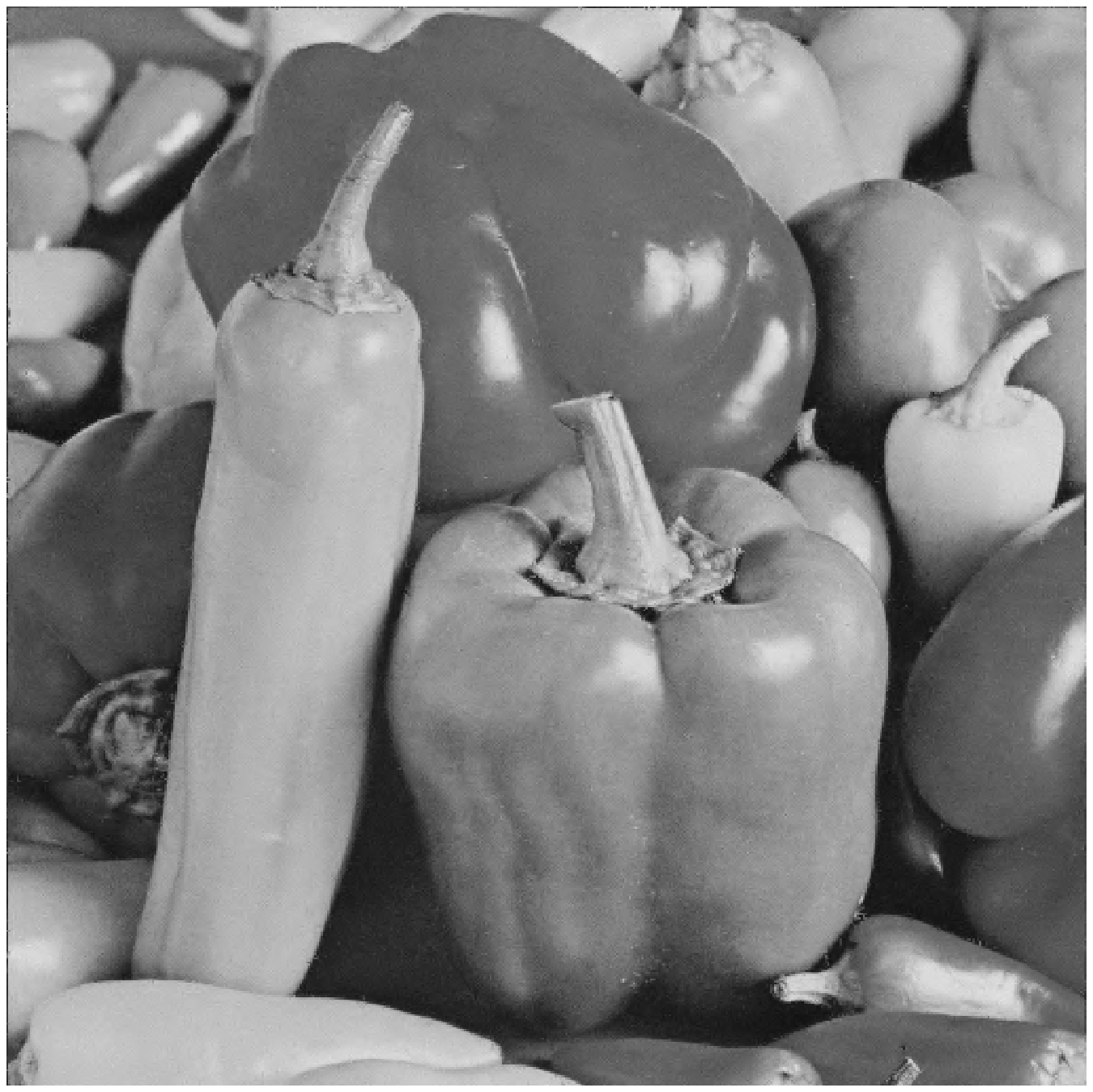}
\end{minipage}}
\subfigure[]{
\begin{minipage}[t]{0.22\linewidth}
\includegraphics[width=\textwidth]{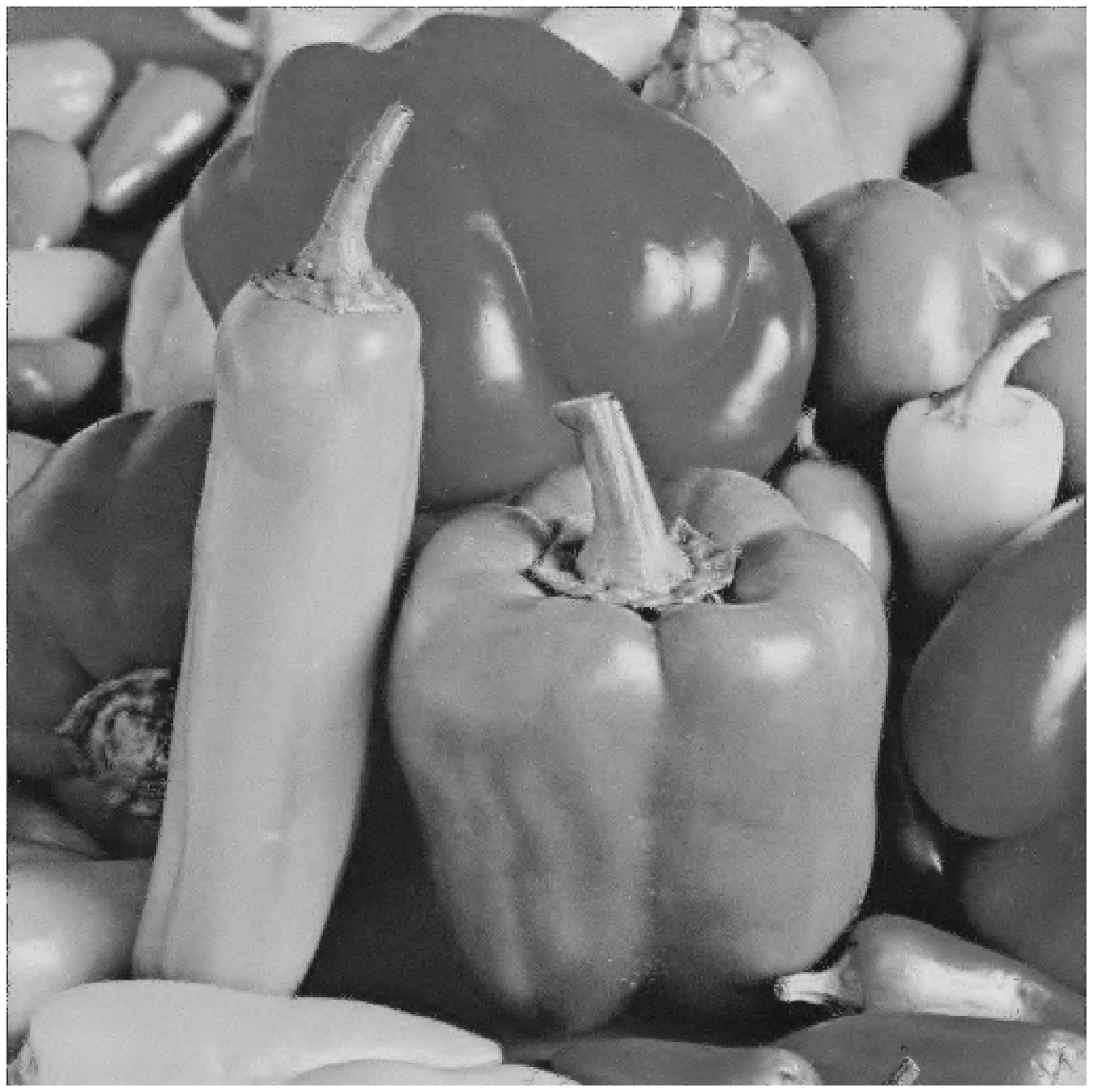}
\end{minipage}}
\caption{\small The reconstructed images and their PSNR values under SR=0.8: (a) PSNR=35.7965, PLR=0.2, BWHT32; (b) PSNR=33.8944, PLR=0.3, BWHT32; (c) PSNR=35.2942, PLR=0.2, BDCT32; (d) PSNR=33.2762, PLR=0.3, BDCT32.}
\label{fig7}
\end{figure*}

\begin{figure*}[th]
\centering
\subfigure[]{
\begin{minipage}[t]{0.22\linewidth}
\includegraphics[width=\textwidth]{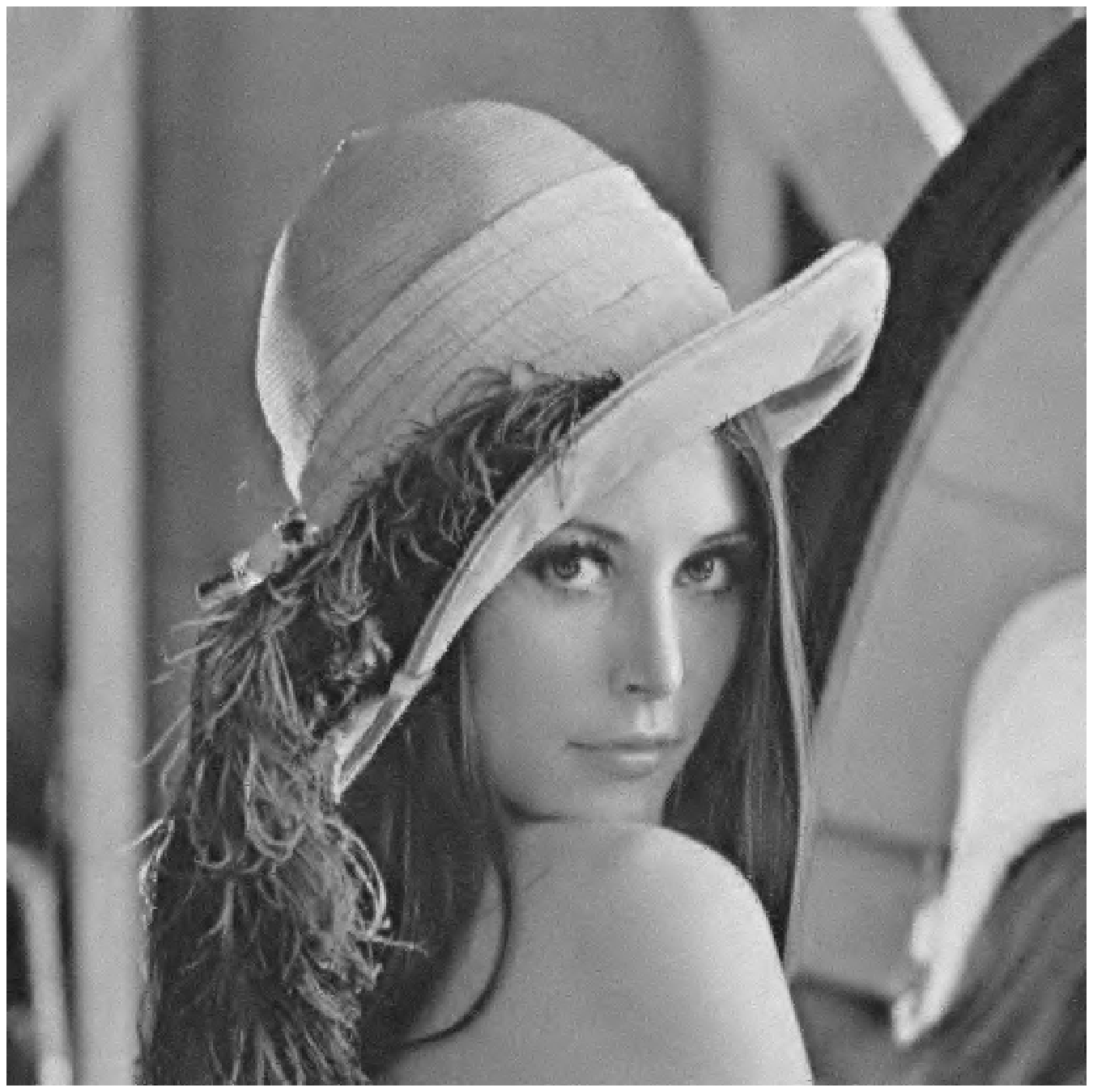}
\end{minipage}}
\subfigure[]{
\begin{minipage}[t]{0.22\linewidth}
\includegraphics[width=\textwidth]{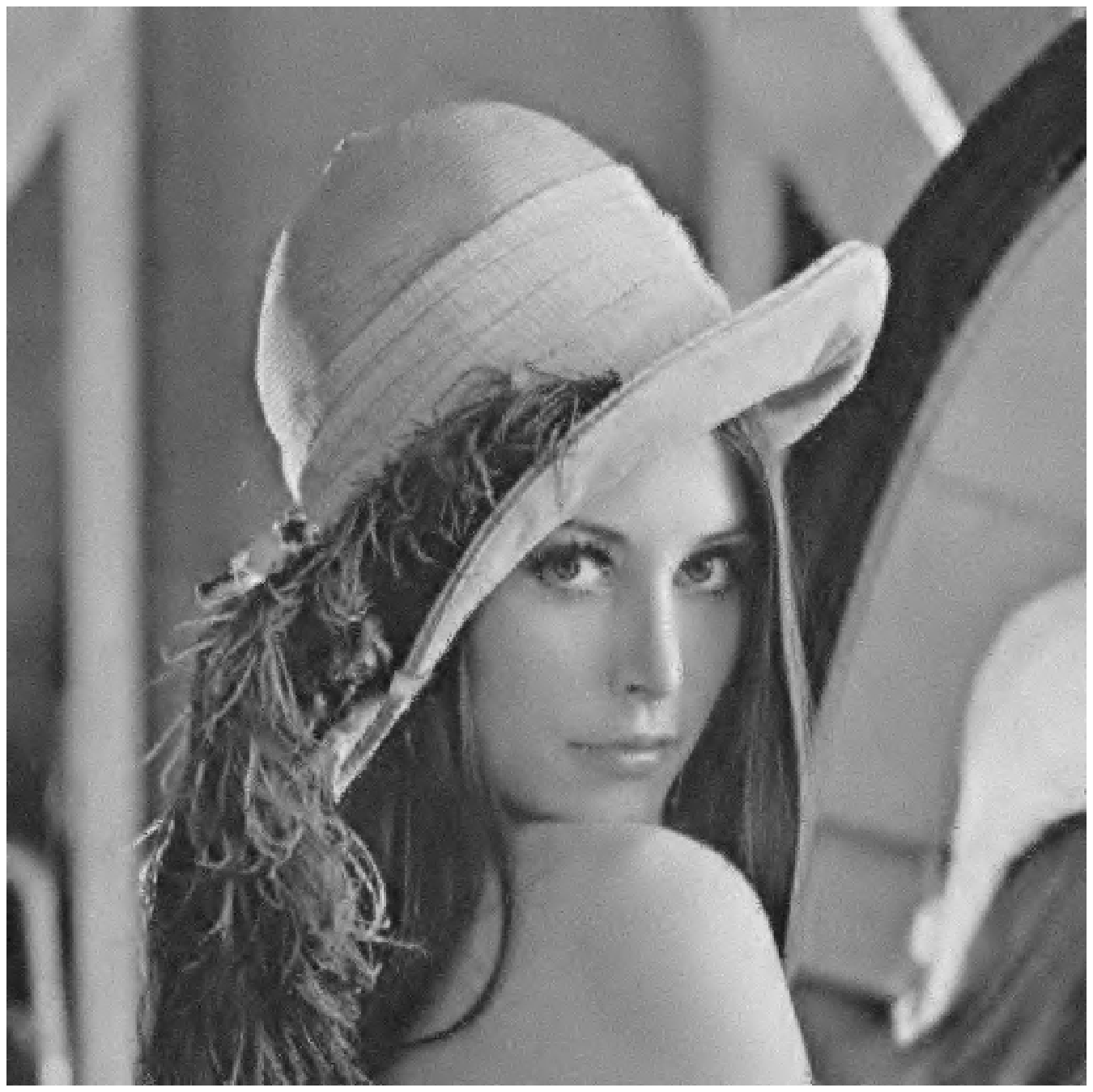}
\end{minipage}}
\subfigure[]{
\begin{minipage}[t]{0.22\linewidth}
\includegraphics[width=\textwidth]{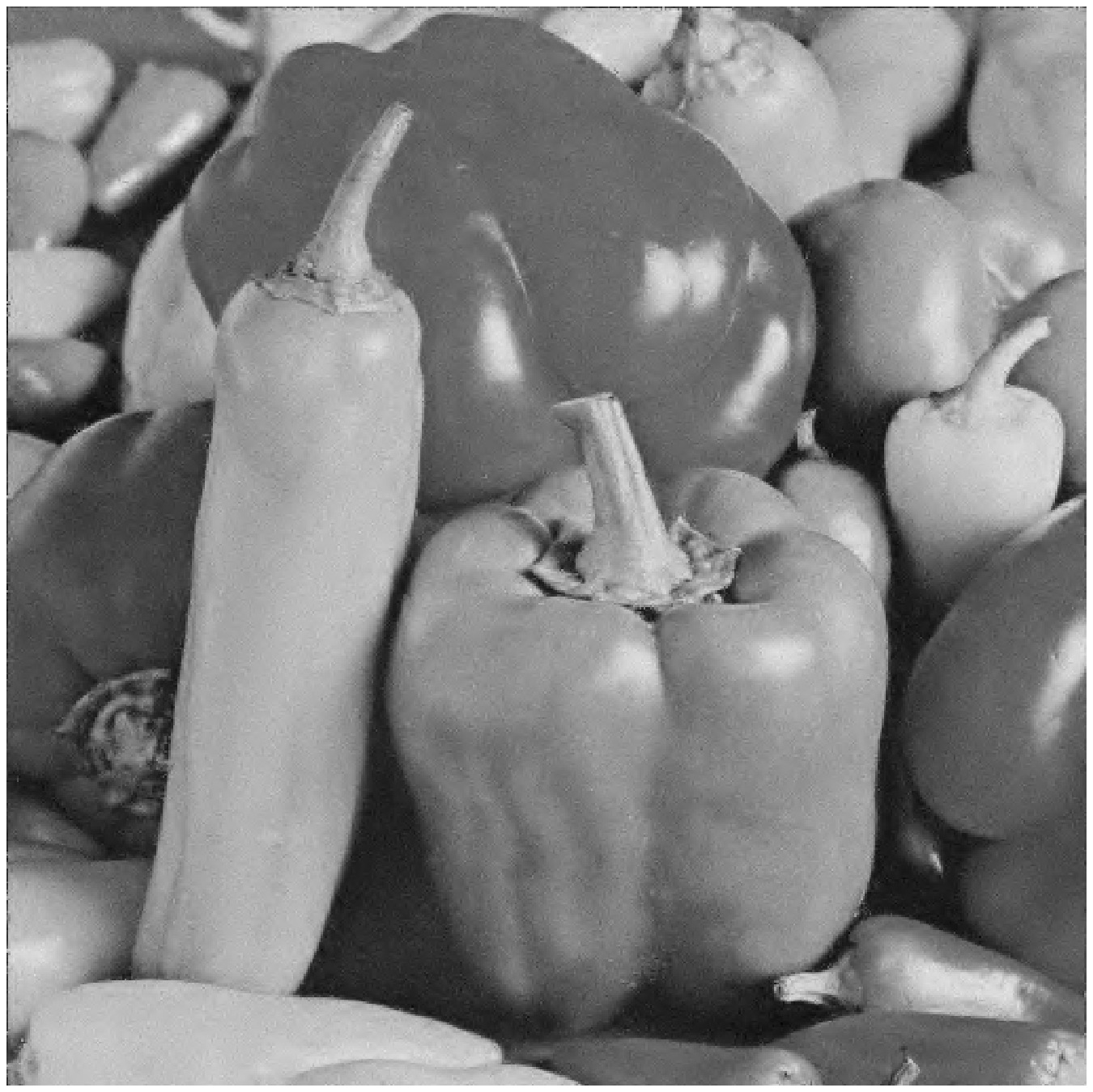}
\end{minipage}}
\subfigure[]{
\begin{minipage}[t]{0.22\linewidth}
\includegraphics[width=\textwidth]{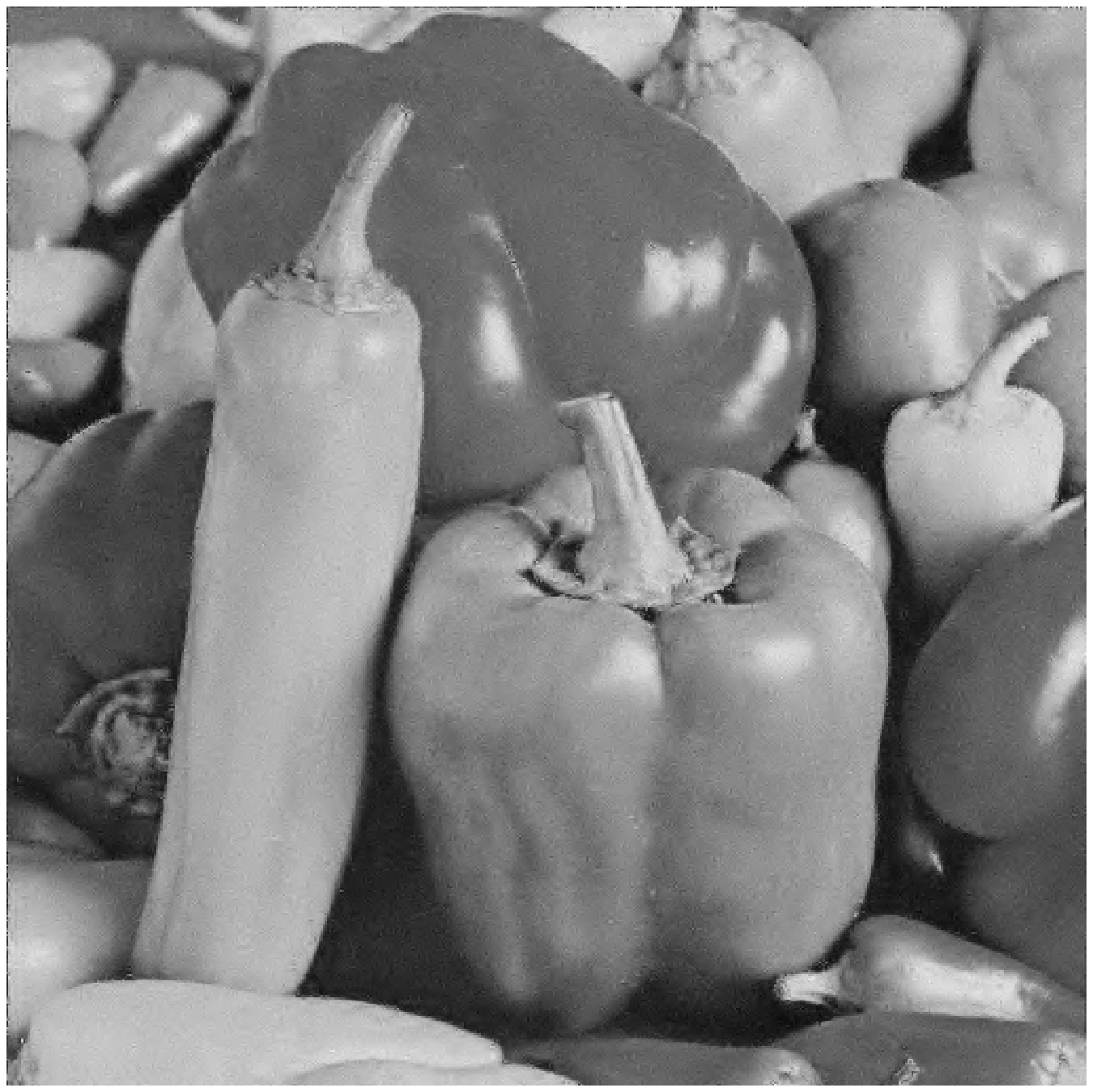}
\end{minipage}}
\caption{\small The reconstructed images and their PSNR values under SR=0.5: (a) PSNR=32.6432, PLR=0.2, BWHT32; (b) PSNR=31.2287, PLR=0.3, BWHT32;
(c) PSNR=32.0629, PLR=0.2, BDCT32; (d) PSNR=30.7208, PLR=0.3, BDCT32.}
\label{fig8}
\end{figure*}

\begin{figure*}[th]
\centering
\subfigure[]{
\begin{minipage}[t]{0.22\linewidth}
\includegraphics[width=\textwidth]{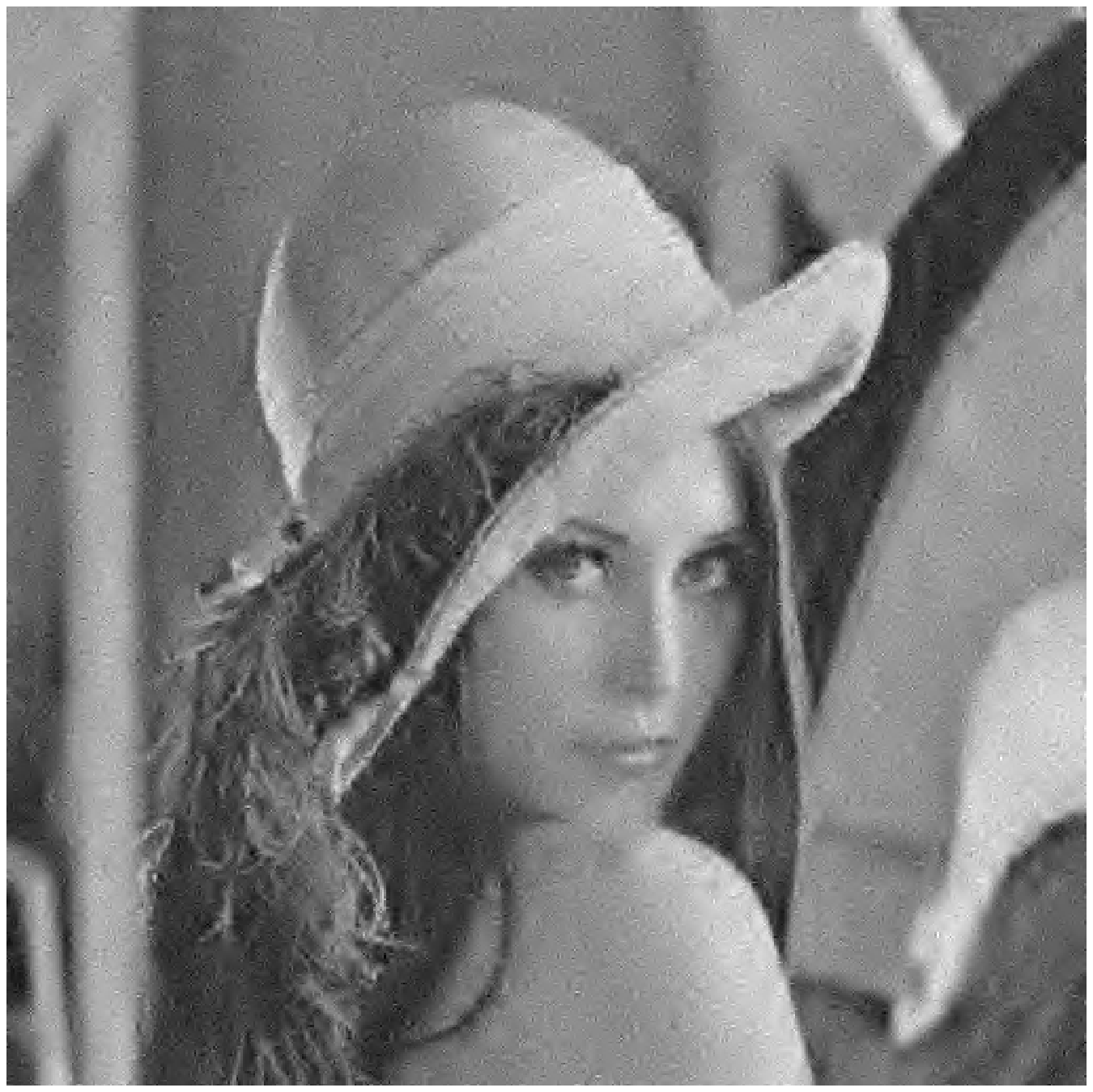}
\end{minipage}}
\subfigure[]{
\begin{minipage}[t]{0.22\linewidth}
\includegraphics[width=\textwidth]{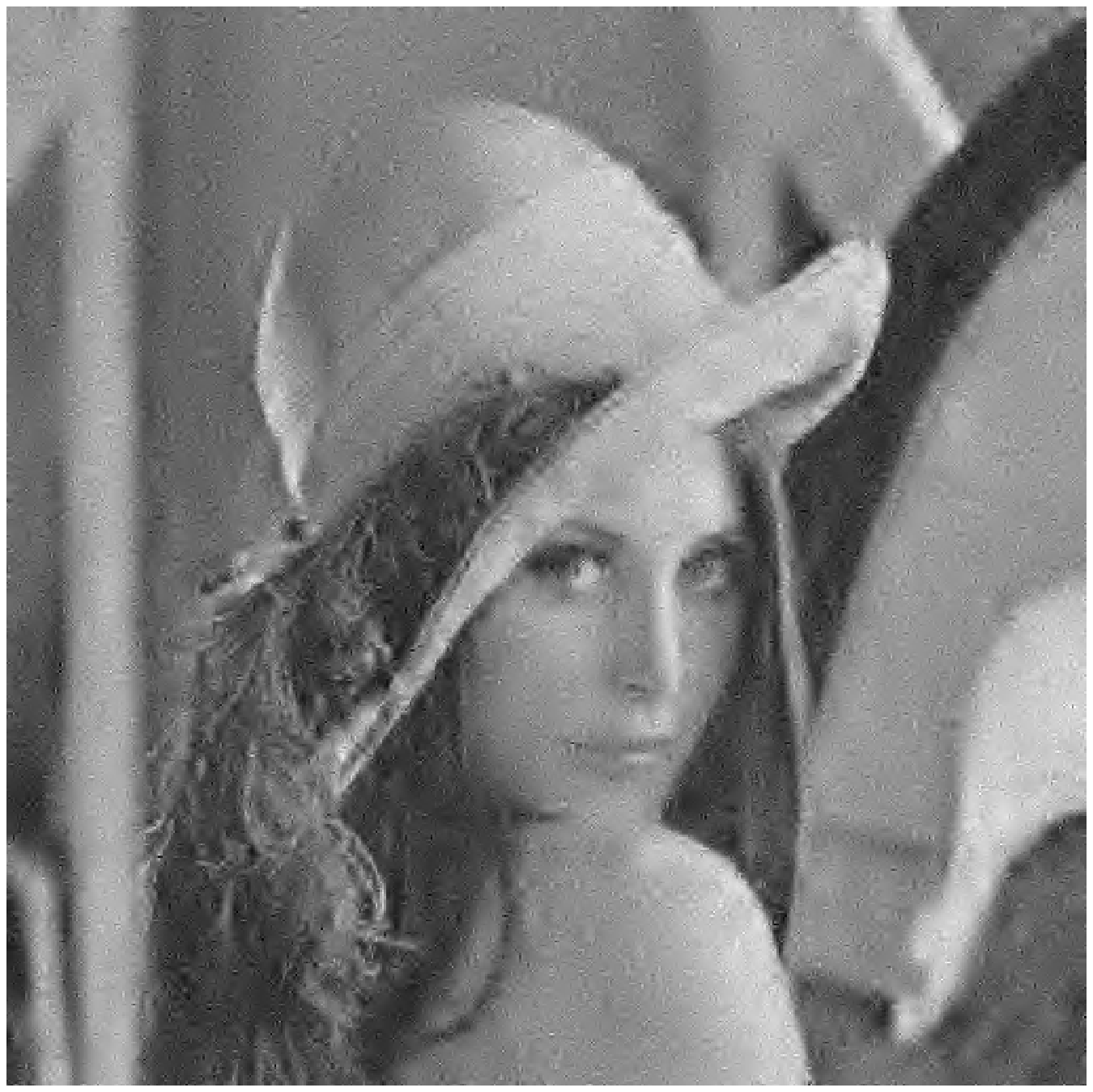}
\end{minipage}}
\subfigure[]{
\begin{minipage}[t]{0.22\linewidth}
\includegraphics[width=\textwidth]{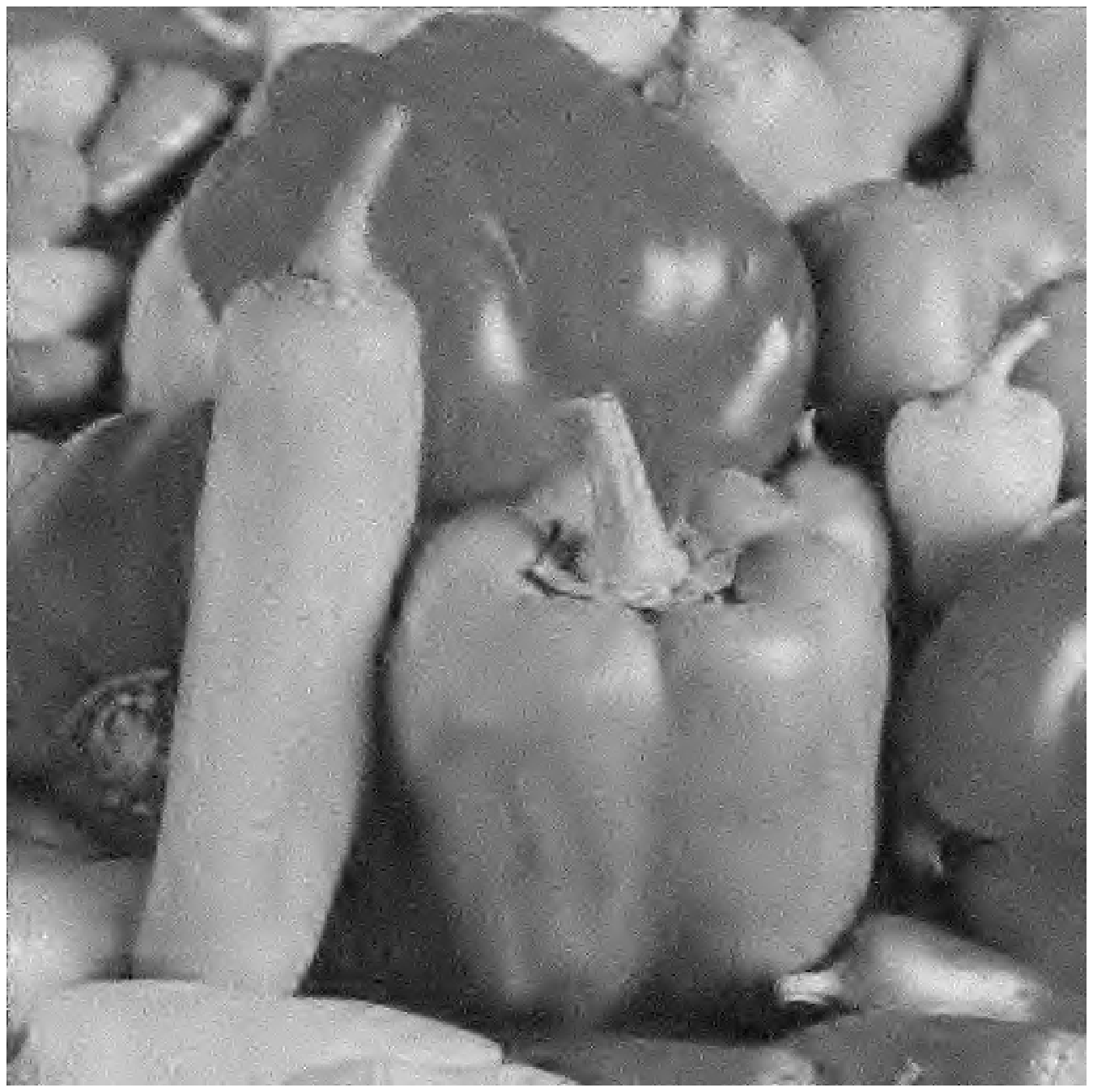}
\end{minipage}}
\subfigure[]{
\begin{minipage}[t]{0.22\linewidth}
\includegraphics[width=\textwidth]{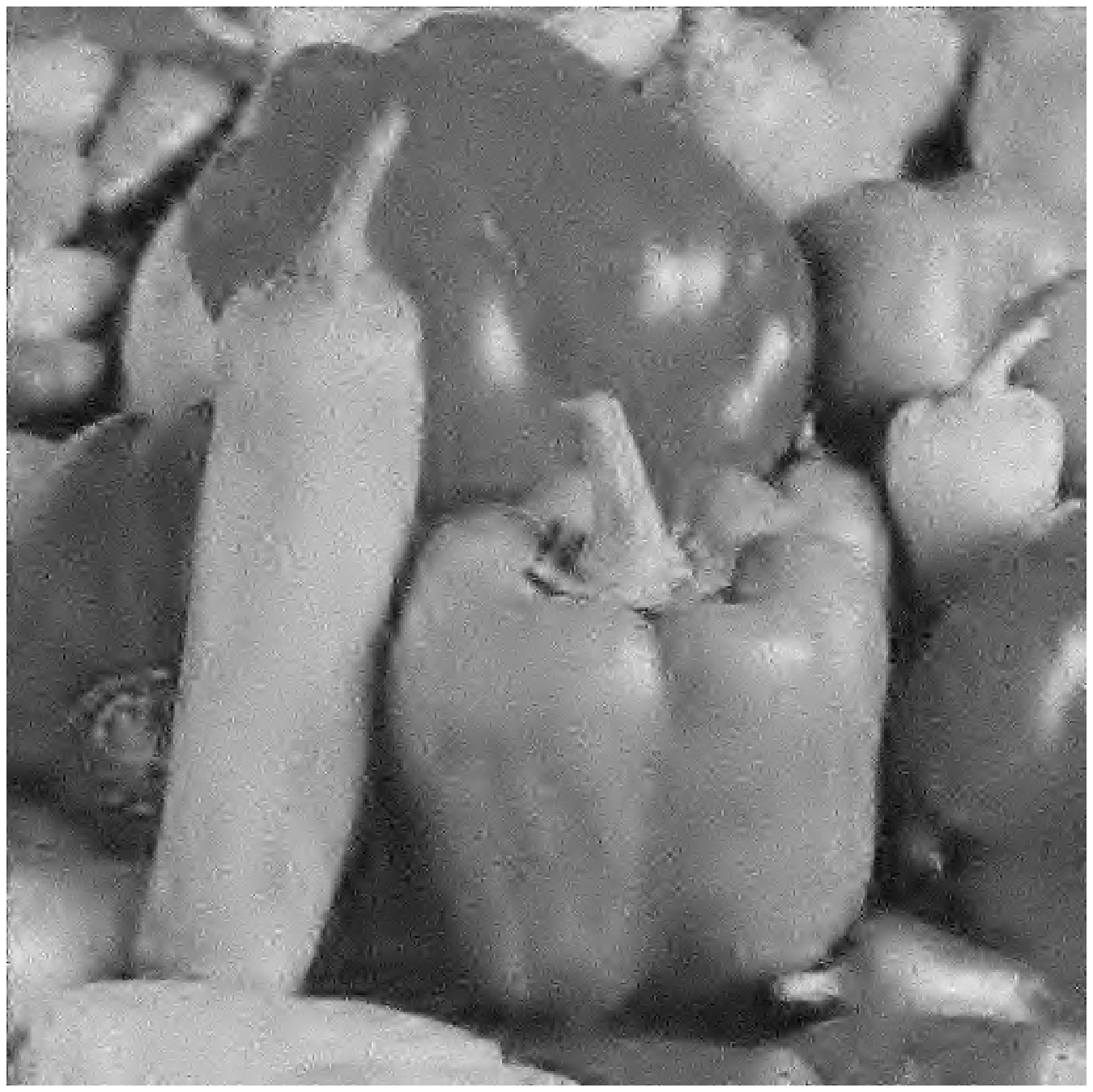}
\end{minipage}}
\caption{\small The reconstructed images and their PSNR values under SR=0.2: (a) PSNR=26.2722, PLR=0.2, BWHT32; (b) PSNR=25.4766, PLR=0.3, BWHT32;
(c) PSNR=26.1835, PLR=0.2, BDCT32; (d) PSNR=24.9015, PLR=0.3, BDCT32.}
\label{fig9}
\end{figure*}

\begin{table*}[tb]
\caption{PSNR versus Round-off and Without Round-off (Lena, SR=0.6, BDCT32).}
\centering
\begin{tabular}{l c c c c c c c}
\hline
$PLR$ & 0 & 0.05 & 0.10 & 0.15 & 0.20 & 0.25 & 0.30 \\
\hline
Round-off & 37.18 & 36.37 &	35.54 &	34.59 &	33.79 & 33.16& 32.15  \\

Without round-off & 37.22&	36.37	&35.66	&34.61	&34.00	&33.25&	32.34  \\

Difference & 0.04 &  0.00  & 0.12 &     0.02 &   0.21 &    0.09  & 0.19  \\

\hline
\end{tabular}
\label{tab1}
\end{table*}

\begin{table*}[tb]
\caption{PSNR versus Round-off and Without Round-off (Lena, SR=0.8, BWHT32).}
\centering
\begin{tabular}{l c c c c c c c}
\hline
$PLR$ & 0 & 0.05 & 0.10 & 0.15 & 0.20 & 0.25 & 0.30 \\
\hline
Round-off & 40.96	&39.27	&38.21&	36.82	&35.70	&34.69	&33.96  \\

Without round-off & 41.01&	39.54&	38.34	&36.82	&35.72&	35.02	&34.00  \\

Difference & 0.05&  0.27 &  0.13   & 0.00   & 0.02  &  0.33  &  0.04 \\

\hline
\end{tabular}
\label{tab2}
\end{table*}

\begin{figure}[th]\centering
 \includegraphics[width=8 cm]{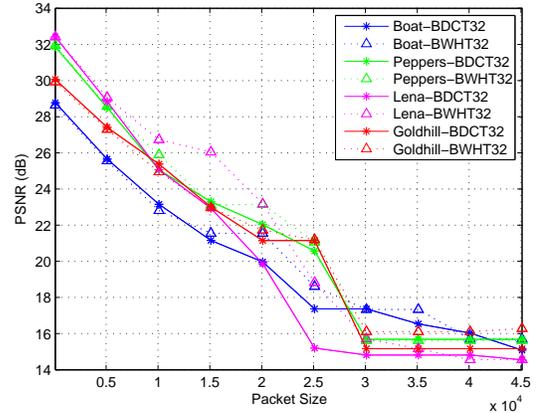}\\
  \caption{PSNR versus packet size when SR=0.6 and PLR=0.3.}
  \label{fig10}
\end{figure}

\begin{figure}[th]\centering
 \includegraphics[width=8 cm]{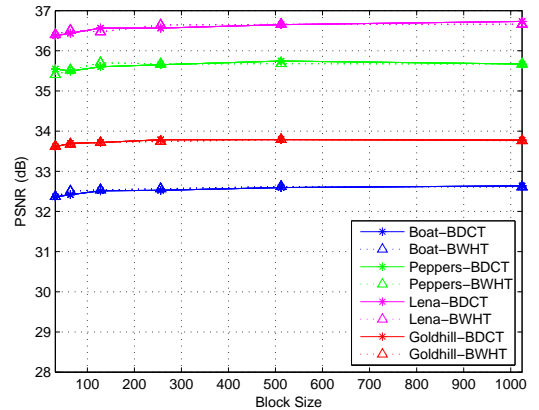}\\
  \caption{PSNR versus block size of SM when SR=0.6 and PLR=0.05.}
  \label{fig11}
\end{figure}

\subsection{Quantization of Compressive Measurements of Cipher Image}

When the compressive measurements are transmitted over a communication channel, they need to be efficiently quantized and encoded. Therefore, the measurements' statistics are required and an optimal quantizer should be tailored to the measurements for minimizing the amount of distortion during reconstruction. The statistical distribution of compressive measurements obtained by SRM has been well studied \cite{haimi2013distribution}. It has been pointed out that the encryption performed by a random permutation on the pixel indices makes the measurements suitable for quantization by causing the measurements' distribution roughly normal. The measurements obtained by applying SM to the encrypted image approximately yield a Gaussian distribution. This is also observed in Fig. 4, which depicts the histograms of various encoded images in different cases.

A uniform scalar quantization is employed to round each entity of ${\bf{y}}$ to the nearest integer. The difference in distortion caused by the round-off is extremely subtle, as shown in Tables I and II. Moreover, we can observe from Fig. 4 that the measurement values roughly lie between -150 and 150. The farther the measurement value deviates from zero, the fewer the number of measurements are required. Our quantization method only reserves and rounds the values located within the interval $[ - 127.5,127.5)$. Others are discarded due to two reasons: (i) The discarded measurements make up only a low proportion, marked as $\gamma $, of the whole measurements. Figure 5 lists the values of $\gamma $ at different parameter settings. $\gamma $ is basically smaller than 0.0055, which implies that either the PLR rises slightly to $PLR = \beta  + \gamma $ or the SR drops a small portion $\alpha \gamma $ by the reason of the approximately equal importance among the measurements; (ii) The reserved measurement values can be one-to-one mapped to the interval $[0,255]$ through adding 128 to every value. The integers in $[0,255]$ not only can be fully represented by 8-bit numbers, but also match with the common-adopted 256 grayscales in the images. After the encoding process is completed, an image can still be stored in 8-bit format, which leads to great convenience in practical usage.

The quantization distortion is caused by two factors: the decimal round-off and the proportion of discarded measurements. The first factor is insignificant, as justified by the data listed in Tables I and II while the second one is the same because $\gamma $ is basically smaller than 0.0055. It can also be justified by the rate-distortion curves plotted in Fig. 6, in which the dashed and solid lines correspond to cases with and without quantization, respectively. These two lines are almost identical and they indicate that the proposed quantization method works well.

\subsection{Robustness}

When the proposed coder is used in a packet network, the robustness is directly related to PLR and SR. Figures 7-9 show some reconstructed Lena and Peppers images at different values of SR and PLR. It can be observed that most of the visual information of the original images can be recovered even when $SR=0.2$ and $PLR=0.3$. This demonstrates that the proposed coder possesses high robustness against packet loss. Besides, the coder does not result in blocking artifacts. In the aforementioned experiments, the packet size is set to 100 while the block size of SM is 32$\times$32. In fact, the robustness is more or less related to both values.

As analyzed previously, there are three factors causing the PSNR difference in exploring the relationship between SR and PLR. Yet these factors arise from the packet size $m$. Intuitively, with an increasing $m$, the PSNR value descends to some extent. This conjecture is justified by Fig. 10, where the parameter settings are $SR=0.6$ and $PLR=0.3$. The smaller the packet size, i.e., the more the number of descriptions, the better the reconstructed image quality is. Naturally, the best case is that each measurement forms a description. When the packet size is between 0 and $3 \times {10^4}$, the PSNR value drops with the reduction in the number of packets. However, when the packet size is larger than $3 \times {10^4}$, the PSNR virtually has no change. This is because that the number of packets is basically reduced to two and remains unchanged. If one of these two packets is lost, it means that half of the successive measurements are sampled. This successional sampling violates the randomness of the down-sampling operator ${\bf{D}}$. The analyses indicate that if the transmission channel allows a small quantity of descriptions and the PLR is too large, for instance, only two descriptions and $PLR \ge 0.3$, the proposed coder cannot be regarded as an efficient multiple description coder. In order to fix this problem, Charlie has to improve the SR. Consider an extreme scenario that $SR=1$, i.e., full redundancy without compression, the encoding process is changed to ${\bf{y = F}}{{\bf{x}}_{en}}$. Such an encoder cannot be guaranteed by the theory of SRM and a great many successive measurements' loss will substantially affect the quality of the reconstructed image. Fortunately, a solution has been developed to cope with this scenario. Associating a realization of down-sampling operator ${\bf{D}}$ that truncates the first or $M$ randomly-selected elements after arbitrarily permuting the signal, Charlie introduces a new random permutation ${\bf{R'}}$ known by Bob. The present encoding form is ${\bf{y = R'F}}{{\bf{x}}_{en}}$. When a packet containing many successive measurements is lost, Bob receives the information ${\bf{\hat y = }}\beta {\bf{R'F}}{{\bf{x}}_{en}}$. Let ${\bf{D'}} = \beta {\bf{R'}}$, which can be considered as a down-sampling operator, then ${\bf{\hat y = D'F}}{{\bf{x}}_{en}}$. In other words, the PLR is the very SR. Even if $PLR = 0.8$, which is equivalent to $SR = 0.8$, the reconstructed image quality is still visually acceptable.

The purpose of having the measurement matrix in a block mode is to reduce storage space and computational complexity at the cost of a lower quality of the recovered signal. In the proposed coder, we investigate PSNR versus the block size of SM when $SR=0.6$ and $PLR=0.05$, as shown in Fig. 11. The greater the block size, the higher the PSNR is. However, the rate of increase is quite slow. Meanwhile, a larger block size of SM needs more memory and consumes more resources. Consequently, a trade-off between them is required. In general, the block size of SM is set as 32$\sim$256.

\section{Conclusion}

A novel and robust coder for processing encrypted images against packet loss has been designed. It is different from the existing approaches of the robust coding of natural images and the compression of encrypted images. The proposed coder based on SRM is composed of three parts: permutation-based encryption by Alice, encoding with structural matrix by Charlie, and joint decryption and decoding by Bob. In addition, we have investigated the relationship between the proposed and the existing methods. Two other cryptographic applications of SRM have also been suggested. In the performance evaluation, we have explored the relationship between packet loss rate and sampling rate. A feasible approach for quantizing the compressive measurements of encrypted images has been introduced. Finally, we have investigated the robustness of the proposed coder at different parameter settings. It has been verified that our coder can be considered as an efficient multiple description coder with a number of descriptions to resist packet loss.

\bibliographystyle{IEEEtr}
\bibliography{ref}

\end{document}